\documentstyle[twoside,fleqn,espcrc2,epsf]{article}
\let\includefigures=\iftrue
\newcommand{\be}{\begin{equation}}
\newcommand{\ee}{\end{equation}}
\newcommand{\beq}{\begin{eqnarray}}
\newcommand{\eeq}{\end{eqnarray}}
\def\ie{{\frenchspacing {\it i.e.}}}
\def\eg{{\frenchspacing {\it e.g.}}}
\def\etal{{\frenchspacing {\it et al.}}}
\def\phicr{\phi_{{\rm cr}}}
\def\omcr{\Omega_{{\rm cr}}}

\newcommand{\AmS}{{\protect\the\textfont2
  A\kern-.1667em\lower.5ex\hbox{M}\kern-.125emS}}

\title{\vbox{
{\baselineskip 12pt
\hbox{\rightline{\rm\small NSF-ITP-94-109}}
\hbox{\rightline{\rm\small hep-th/9411020}}
}
\hbox{Black Hole Evolution}}}

\author{L\'arus Thorlacius\thanks{Supported in part by NSF grant
No. PHY89-04035.}\\
\ \\
Institute For Theoretical Physics, University of California\\
Santa Barbara, CA 93106-4030, USA}%

\begin{document}

\begin{abstract}
Black hole formation and evaporation is studied in the
semiclassical approximation in simple 1+1-dimensional
models, with emphasis on issues related to Hawking's
information paradox.  Exact semiclassical solutions are
described and questions of boundary conditions and vacuum
stability are discussed.

The validity of the semiclassical approximation has been
called into question in the context of the information
puzzle.  A different approach, where black hole
evolution is assumed to be unitary, is described.
It requires unusual causal properties and kinematic
behavior of matter that may be realized in string
theory.

Based on lectures given at the 1994 Trieste Spring School
on String Theory, Gauge Theory, and Quantum Gravity.
\end{abstract}

\maketitle

\setcounter{footnote}{0}
\section{Introduction}
\label{intro}

Hawking's discovery of black hole evaporation \cite{hawking1} has
presented us with a unique window on the interplay between gravity
and quantum physics.
The puzzles of black hole evolution highlight
the incompatibility between the world view offered by Einstein's
general relativity and that of quantum theories.  The so called
black hole information paradox, first formulated by Hawking
\cite{hawking2}, is a good example.

Consider an initial state of diffuse matter undergoing gravitational
collapse and assume that the system starts out in a pure quantum
mechanical state.  According to classical gravitational theory,
once the collapsing matter approaches its own
Schwarzschild radius the external world only retains information
about the total energy and conserved gauge charges carried by the
infalling matter.  This follows from a series of uniqueness theorems
for solutions of general relativity coupled to various types of matter
fields \cite{hawell}, which are collectively referred to as
``no-hair'' theorems.
Observers outside the black hole no longer have access to all
the degrees of freedom of the system and cannot describe it by a pure
state.  There is nothing wrong with that, as long as we consider only
classical solutions of the gravitational field, for we can postulate
that the missing information is hidden in the black hole and that the
quantum state of the whole system remains pure even if outside observers
have to resort to a density matrix description.

But now the black hole evaporates and according to Hawking's
semiclassical calculation the emitted radiation is thermal.  Let us
for the moment assume that the evaporation leaves no remnant behind
so that the final state consists only of outgoing Hawking radiation,
which, being thermal, is described by a mixed quantum state.
It then appears that this process of black hole formation and
subsequent evaporation evolves a pure state into a mixed one in direct
violation of the basic rule that every quantum mechanical system should
have a unitary evolution.

Hawking suggested to take this at face value and claimed that this
is an example of an added fundamental uncertainty introduced into
quantum physics, over and above the usual Heisenberg uncertainty,
when gravitational effects are taken into account \cite{hawking3}.
He further proposed a modified set of axioms for quantum field theory,
which would allow pure states to evolve into mixed states.
In his formalism the unitary $S$-matrix of quantum field theory, which
maps an initial quantum state to a final state, is to be replaced by a
superscattering operator {\sl \$}, which maps an initial density matrix
to a final density matrix.  In ordinary quantum field theory the
superscattering operator factorizes, {\sl \$}~$=S \bar S$, but when
gravity enters the game this is no longer true, due to black hole
formation or even virtual processes involving gravitational
fluctuations.  Hawking's proposal was criticized by a number of
authors \cite{ehns,bsp,gross}, and
recently Strominger \cite{strominger} pointed out that Hawking's
original proposal violates the superposition principle.

An alternative viewpoint, first suggested by Page \cite{page1} and
strongly advocated by 't~Hooft \cite{thooft1}, is that the Hawking
radiation is not really exactly thermal but in fact carries all the
information about the initial state of the infalling matter.
This information is encoded in subtle correlations between quanta
emitted at different times during the evaporation process, and
detecting it would require a large number of statistical observations
to be made on an ensemble of identically prepared states.  This
viewpoint is a conservative one from the point of view of quantum
theory, since it insists on the existence of a unitary $S$-matrix,
but it appears to require radical assumptions about the kinematic
behavior of matter at high energies \cite{thooft1,susskind}.
Sections~\ref{secv}-\ref{secvii} of these lecture notes will be
concerned with some recent work where this viewpoint is adopted.

A third possibility, advocated by Aharonov, Casher,
and Nussinov \cite{acn}, is that a black hole does not
completely evaporate and the information
is carried off by a Planck scale remnant.  There would need to be a
distinct remnant for each possible initial state, so the density of
these remnant states at the Planck energy must be virtually infinite.
This leads to thorny phenomenological problems if the remnants
behave at all like local objects and their effects on low-energy
physics can be described by an effective field theory.  Even if
individual remnant states couple extremely weakly to processes such
as $e^+$-$e^-$ scattering at a colliding beam facility, the infinite
density of states would cause them to be the dominant channel.  One
would also expect a divergent pair production rate of remnants in
weak background fields and thermal sums would be rendered ill defined
by their contribution.  None of these effects are observed so
either black holes do not leave behind information carrying remnants
or those remnants are described by unconventional laws of physics at
low energies.  Considerable effort has gone into developing a picture
of remnant dynamics \cite{remnants} where the above pathologies are
to be avoided.
This work was described in detail in the lectures of Banks at this
Spring School.

Recently Polchinski and Strominger \cite{jpas} have argued that
Hawking's superscattering approach can be successfully reformulated
as a unitary theory, with long-lived remnants, in the context of a third
quantized theory of gravity.  This interesting proposal will not
be discussed here.

The information puzzle is an important theoretical
problem because the resolution of
the paradox may require a revision of some fundamental physical laws.
Since we are unlikely to obtain laboratory data on
quantum effects in gravity any time soon, confronting our theories
with physical paradoxes of this type is one of the most promising
lines of inquiry in this area of theoretical physics.

These lecture notes consist of two parts, each of which is to a
large extent self-contained.
The first part is a review of black hole physics in two spacetime
dimensions with a view towards the information problem.
In Section~\ref{secii} we introduce a two-dimensional model, proposed
by Callan, Giddings, Harvey, and Strominger (CGHS) \cite{cghs}
for black hole physics, and study its
classical solutions.  In Section~\ref{seciii} we consider the
quantum theory of matter fields in a classical background black hole
geometry and exhibit the two-dimensional analog of the Hawking effect.
We give simple arguments for the thermal character of the Hawking
radiation and then show how its back-reaction on the geometry can
be accounted for via a set of semiclassical corrections
to the equations of motion.  In Section~\ref{seciv} we adopt a more
systematic approach to the quantization of these two-dimensional
models.  We consider conformally invariant effective theories which
reduce to the CGHS model in the classical limit.
We address a number of issues which come up in the study of these
models, such as the rate of Hawking evaporation,
boundary conditions in the strong coupling region, and vacuum
stability.

The second part of the lecture notes, beginning with
Section~\ref{secv}, describes a different approach to black hole
evolution where it is assumed from the outset that the information
is returned in the Hawking radiation.  In Section~\ref{secv} we
put forward a phenomenological framework for black hole physics
which is consistent with a unitary evolution of quantum states.
It is argued that any model where information is returned encoded
in the Hawking radiation will have to incorporate a principle of
black hole complementarity, which allows for the different
viewpoints of an observer, who enters a black hole in free fall,
and of an observer who remains outside at all times.  In
Section~\ref{secvi} we consider some gedanken experiments designed
to test the validity of the complementarity hypothesis and find
that their detailed analysis requires knowledge of Planck scale
effects.  This indicates that the information paradox is not
well posed in terms of low-energy physics alone.  In
Section~\ref{secvii} we describe some recent work which suggests
that string theory implements black hole complementarity in a
natural way.  The key observation in this context is that
string matter exhibits very different kinematic behavior
at high energies than matter formed out of weakly interacting
pointlike particles.

\section{Classical Dilaton Gravity in 1+1 Dimensions}
\label{secii}
When faced with a difficult problem it is often useful
to look for a simpler toy system, in which an analogous problem can
be posed and studied and, in the best of all worlds, solved.
In the case of the black hole information puzzle such a simplified
context is
provided by certain two-dimensional models of gravity which have
been actively studied (but unfortunately not fully solved)
in recent years.  These theories are far from
being realistic models of real gravity since crucial ingredients of
the four-dimensional physics, such as propagating gravitons, are
missing.  The simple toy theories do, however,
have black hole geometries as classical solutions.
When one considers the quantum
theory of matter fields in such spacetimes one finds Hawking radiation
and, at the semiclassical level, its back-reaction on the geometry
leads to an information paradox, which is entirely analogous to the
one posed by Hawking.  The fate of quantum information is an important
question of principle and it seems worth looking for an answer in
this simplified context even if it is not at all guaranteed to reflect
accurately on the situation in a more realistic setting.

A large number of papers has been written on various aspects of
two-dimensional black hole physics in recent years.  For reviews
see \eg\ \cite{jhas,giddings}.

\subsection{The CGHS Model}
\label{twoone}
The CGHS model \cite{cghs} of two-dimensional dilaton gravity,
coupled to scalar matter fields, was proposed a few years ago
as a particularly convenient toy model for black hole physics.
The classical dynamics is governed by the action
\beq
\lefteqn{S_{0}=\frac{1}{2\pi}\int d^2y \sqrt{-g}
\Bigl[e^{-2\phi} (R+4(\nabla\phi)^2 +4\lambda^2)} \nonumber \\
& &
- \frac{1}{2} \sum^N_{i=1} (\nabla f_i)^2 \Bigr] \,,
\label{classact}
\eeq
which can be viewed as an effective action for radial modes of
near-extremal magnetically charged black holes in four-dimensional
dilaton gravity \cite{cghs,gs1,bddo}.  We will primarily be
interested in this theory on its own merits as a two-dimensional model
of gravity coupled to matter, but the higher-dimensional
interpretation is helpful in developing an intuitive
picture of some aspects of the physics.

The action (\ref{classact})
inherits a length scale $\lambda^{-1}$ from the four-dimensional
geometry, which is set by the magnetic charge of the extremal
black hole, $\lambda^{-1}=2Q$.  We shall use units in which $\lambda=1$
throughout.
In the region of the four-dimensional geometry where the two-dimensional
effective description applies, the physical radius\footnote{
This is the radius measured by the Einstein metric.  If we instead use
the string metric the radius would be constant in this region,
which is accordingly often referred to as the `infinite throat'
part of the four-dimensional geometry.  See \cite{horowitz} for a
detailed discussion of black holes in four-dimensional dilaton gravity.}
of the local transverse two-sphere is given by the dilaton field,
$r(x^0,x^1)=e^{-\phi(x^0,x^1)}$.

The classical equations of motion are
\be
\!\!\begin{array}{l}
\nabla^2 f_i = 0\,, \\
{  } \\
{1\over 4}R+\nabla^2\phi-(\nabla\phi)^2+1 = 0\,, \\
{  } \\
\nabla_\mu\nabla_\nu\phi+g_{\mu\nu}((\nabla\phi)^2-\nabla^2\phi-1)
={e^{2\phi}\over 2} T^f_{\mu\nu}\,,
\end{array}
\label{classeqs}
\ee
where $T^f_{\mu\nu}$ is the matter energy-momentum tensor,
\be
T^f_{\mu\nu} = {1\over 2}\sum_{i=1}^N
\bigl[\nabla_\mu f_i \nabla_\nu f_i
- {1\over 2}(\nabla f_i)^2\bigr]\,.
\label{tmatter}
\ee
It is convenient to work in conformal gauge and choose lightcone
coordinates $y^\pm=y^0\pm y^1$, for which the line element is
\be
ds^2=-{1\over 2}e^{2\rho}dy^+dy^- \,.
\label{lineel}
\ee
The fields in the theory are then $f_i$, $\phi$, and the conformal
factor $\rho$.  The classical equations of motion of these fields can
be arranged to read
\be
\!\!\begin{array}{l}
\partial_+\partial_-f_i = 0\,, \\ { } \\
\partial_+\partial_-(e^{-2\phi}) = -e^{2(\rho-\phi)}\,, \\ { } \\
\partial_+\partial_-(\rho-\phi) = 0 \,.
\end{array}
\label{confeqs}
\ee
In addition, one must impose as constraints the equations of motion
corresponding to the components of the metric that have been set
to zero by this choice of gauge,
\be
\!\!\begin{array}{l}
e^{-2\phi}(2\partial_+^2\phi-4\partial_+\rho\partial_+\phi) =
T^f_{++} \,, \\ { } \\
e^{-2\phi}(2\partial_-^2\phi-4\partial_-\rho\partial_-\phi) =
T^f_{--} \,.
\end{array}
\label{constraints}
\ee
The non-vanishing components of the matter energy-momentum tensor
are given by
$T^f_{\pm\pm}={1\over 2}\sum_{i=1}^N (\partial_\pm f_i)^2$,
and the conservation of matter energy-momentum
takes the form
$\partial_-T^f_{++}=0=\partial_+T^f_{--}$ in the classical theory.
The left moving
energy flux $T^f_{++}$ is only a function of the left moving lightcone
coordinate $y^+$ and the right moving flux $T^f_{--}$ only depends on
$y^-$.

One of the conformal gauge equations (\ref{confeqs}) is
$\partial_+\partial_-(\rho-\phi)=0$, which has the general solution
$\rho=\phi+f_+(y^+)+f_-(y^-)$.  The arbitrary functions $f_+$ and
$f_-$ can be eliminated by a conformal reparametrization to
coordinates $(x^+,x^-)$ such that
$dx^+/dy^+=e^{2f_+}$ and $dx^-/dy^-=e^{2f_-}$.
This special coordinate system is referred to as {\it Kruskal
coordinates\/} for reasons which will become apparent a little
later on.

Since $\rho=\phi$ in
Kruskal gauge the equations of motion and constraints
reduce to
\be
\!\!\begin{array}{l}
\partial_+\partial_-e^{-2\phi}=-1 \,, \\ { } \\
\partial_\pm^2e^{-2\phi(x^+,x^-)}
=- T^f_{\pm\pm}(x^\pm) \,,
\end{array}
\label{kruskaleqs}
\ee
Let us first consider solutions with vanishing flux of matter energy,
$T^f_{\pm\pm}=0$.  The simplest one is the so called {\it linear
dilaton vacuum\/},
\be
\!\!\begin{array}{l}
f_i = 0 \,, \\ { } \\
e^{-2\phi} = e^{-2\rho} = -x^+x^- \,.
\end{array}
\label{ldv}
\ee
This geometry has vanishing curvature everywhere.  It derives its
name from its expression in the coordinate system $(\sigma^+,\sigma^-)$
defined by the transformation $x^\pm=\pm e^{\pm\sigma^\pm}$, where
the metric is manifestly flat, $\rho=0$, and the dilaton is linear in
the spatial coordinate, $\phi=-(\sigma^+-\sigma^-)/2=-\sigma^1$.

Due to the factor of $e^{-2\phi}$ in front of the dilaton-gravity
terms in the action (\ref{classact}) the value of the dilaton field
controls the strength of gravitational quantum corrections in the
theory.  In the linear dilaton vacuum the coupling varies
with spatial position, ranging monotonically from infinite strength
in the limit $\sigma^1\rightarrow -\infty$ to zero as
$\sigma^1\rightarrow +\infty$.  From the 3+1-dimensional viewpoint
$\sigma^1$ is a radial coordinate and the weakly coupled region
corresponds to asymptotic transverse two-spheres of large radius, while
the strong coupling at $\sigma^1\rightarrow -\infty$ reflects the fact
that the transverse area is going to zero and short-distance effects
are becoming important.
In general one expects significant quantum corrections to the
spacetime metric where the coupling is strong and in some models the
internal asymptotic region is replaced, as we shall see later on, by
a timelike boundary which can be interpreted as the origin of radial
coordinates.

\subsection{Eternal Black Holes}
\label{twotwo}
In classical general relativity
a {\it black hole\/} is defined as a region of spacetime which is not
in the causal past of future null infinity ${\cal I}^+$ \cite{hawell}.
This means that no timelike observer can escape from a black hole since
even null radiation is trapped.  The boundary of the black hole is a
null surface, called the {\it global event horizon}.
Local observers cannot determine from local
initial data whether they are inside a black hole.  In order to locate
the global event horizon one must have knowledge of future evolution of
the entire spacetime manifold and be able to find the causal past of
${\cal I}^+$.

The linear dilaton vacuum (\ref{ldv}) is a special case of a
one-parameter
family of static solutions:
\be
\!\!\begin{array}{l}
f_i = 0 \,, \\ { } \\
e^{-2\phi} = e^{-2\rho} = M_0-x^+x^- \,.
\end{array}
\label{staticbh}
\ee
The scalar curvature of these geometries is
\be
R=8e^{-2\rho}\partial_+\partial_-\rho={4M_0\over M_0 - x^+x^-} \,.
\label{staticcurv}
\ee
For $M_0\neq 0$ there are
two curvature singularities which asymptotically approach the null
curves $x^\pm=0$.  The gravitational coupling strength diverges at these
singularities.

If $M_0<0$ the singularities are timelike and the future of any Cauchy
surface contains a naked singularity, {\frenchspacing {\it i.e.}}
one which is visible from ${\cal I}^+$.

For $M_0>0$ the curvature singularities are spacelike.
One of them is a white hole
singularity, which is not in the causal future of any event, and the
other one is a black hole singularity encompassed by a global event
horizon, which consists of two null line segments,
$\{x^+{=}0,\ x^-{>}0\}$ and $\{x^-{=}0,\ x^+{>}0\}$.
The parameter $M_0$ is proportional to the canonical ADM mass of the
black hole \cite{witten}.

\begin{figure}[bt]
\epsfxsize=7.cm \epsfysize=6cm
\epsfbox{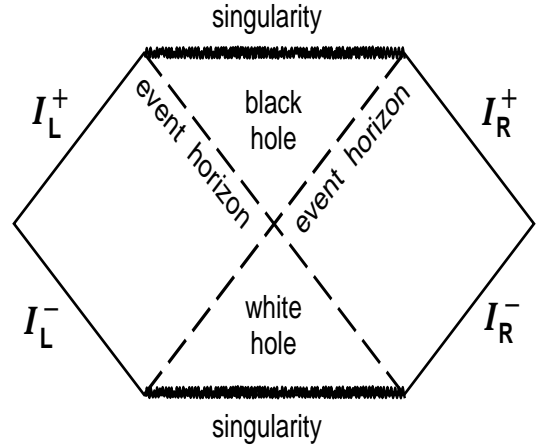}
\caption{Penrose diagram for static black hole.}
\label{sbhpenr}
\end{figure}

The Penrose diagram in Figure~\ref{sbhpenr} is obtained by making the
conformal reparametrization $q^\pm = \arctan{(x^\pm/\sqrt{M_0})}$.
It shows that the global causal structure of a static
1+1-dimensional black hole is completely analogous to that of the
maximally extended Schwarzschild solution in
3+1~dimensions\cite{hawell}.

\includefigures\fi

The value of the dilaton field at the event horizon is
$e^{2\phi_H}=1/M_0$,
so the strength of the gravitational coupling at the event horizon
can be made arbitrarily weak by considering a black hole of large
mass.  The scalar curvature at the event horizon is
$R_H=4$, independent of the black hole mass.

In the asymptotic region, where $x^+x^-\rightarrow -\infty$, the
curvature
goes to zero.  The coordinate transformation
$\sigma^\pm=\pm\log{x^\pm}$,
puts the metric into a manifestly asymptotically flat form:
\be
e^{2\rho} = (1+M_0\,e^{-2\sigma^1})^{-1} \,.
\label{tortoiselineel}
\ee
The $\sigma$-coordinate system only covers the region of spacetime
which is outside the global event horizon and is
analogous to the tortoise coordinates used to describe the
Schwarzschild solution outside a black hole in 3+1 dimensions.
The maximal analytic extension of the Schwarzschild solution is
obtained by transforming to Kruskal coordinates and the same is
achieved in the 1+1-dimensional case by going to the $(x^+,x^-)$
coordinate system, which is named accordingly.

We close the discussion of static solutions of the classical CGHS model
by considering Euclidean two-dimensional black holes.
The tortoise time coordinate $\sigma^0$ can be analytically continued
to a Euclidean time coordinate $\theta$ and if we also make a
spatial reparametrization:
$\exp{\sigma^1}=\sqrt{M_0}\sinh{\rho}$,
the line element takes the form
\be
ds^2=d\rho^2+{\sinh ^2 \rho \over 1 + \sinh ^2 \rho}\,d\theta^2 \,.
\label{infcig}
\ee
Near the global event horizon at $\rho=0$ the Euclidean coordinates
reduce to standard polar coordinates, with angle $0<\theta<2\pi$,
but in the asymptotic region, where $\rho\rightarrow\infty$, they
parametrize a flat cylinder.
We would like to interpret the Euclidean solution in terms of a black
hole in thermal equilibrium with a gas of matter, and infer the
equilibrium temperature, $T=1/2\pi$, from the periodicity of $\theta$.
Although this indeed gives the correct Hawking temperature for
two-dimensional black holes in this theory, such an interpretation is
premature, especially in light of the fact that any static solution
of the classical equations (\ref{confeqs}) and (\ref{constraints})
with non-vanishing matter energy density outside a black
hole has a singular event horizon.  As we shall
see in Section~\ref{seciii}, this issue gets resolved at the
semiclassical level, where we include the back-reaction on the
geometry due to Hawking radiation, and find that a black hole in
equilibrium with a heat bath at the Hawking temperature
is described by a smooth geometry.

\subsection{Classical Gravitational Collapse}
\label{twothree}

We now turn our attention to dynamical solutions of the classical
equations of motion.  In general these can have both outgoing and
incoming energy flux but the geometries of most physical interest
describe black hole formation from the vacuum by leftmoving matter,
\be
\!\!\begin{array}{l}
f_i = f_i^+(x^+) \,, \\ { } \\
e^{-2\phi}=e^{-2\rho} = M(x^+)-x^+\bigl(x^-{+}P_+(x^+)\bigr) .
\end{array}
\label{collapse}
\ee
The infalling matter only influences the geometry
through two moments of the incoming energy flux.
\be
M(x^+) = \int_0^{x^+} dy^+ \, y^+\,T^f_{++}(y^+) \,,
\label{energy}
\ee
is the matter energy incident from ${\cal I}^+_R$ before advanced
time $x^+$, and
\be
P_+(x^+) = \int_0^{x^+} dy^+ \, T^f_{++}(y^+) \,,
\label{kruskalmomentum}
\ee
is referred to as the Kruskal
momentum of the incoming matter distribution.
The functions $f_i^+$ can be quite general as long as the total incoming
energy, $M_\infty =M(x^+{=}\infty)$, and Kruskal momentum,
$P_\infty =P_+(x^+{=}\infty)$, are finite.

Consider an incoming
matter flux which is switched on for a finite time interval, \ie\
the functions $f_i^+(x^+)$ are taken to be nonvanishing only on some
interval $x^+\in [x^+_1,x^+_2]$.
At early advanced times, $x^+<x^+_1$, the solution (\ref{collapse})
then reduces to the linear dilaton vacuum (\ref{ldv}) while at late
advanced times, $x^+>x^+_2$, it takes the form of an eternal black
hole (\ref{staticbh}) with $M_0$ replaced by $M_\infty$ and the
$x^-$ coordinate shifted by the total Kruskal momentum $P_\infty$.

The scalar curvature of the dynamical solution (\ref{collapse}) is
\be
R={4M(x^+) \over M(x^+)-x^+(x^-+P_+(x^+))} \, .
\label{curv}
\ee
There is a spacelike black hole singularity on the contour
\be
x^-_S(x^+)=-P_+(x^+)+M(x^+)/x^+\,,
\label{sing}
\ee
which is asymptotic to the null line
$x^-=-P_\infty$ in the $x^+\rightarrow\infty$ limit.  This null line
defines the global event horizon of the black hole.  As expected,
its location can only be determined at the end of the day because
all the incoming energy flux contributes to $P_\infty$.

Contours of constant $\phi$ are spacelike in the region near the
black hole singularity.  Locally the area of the transverse
two-sphere decreases along both future null directions
there, which means it is a region of future trapped points.
The outer boundary of the trapped region defines the {\it apparent
horizon\/} of the black hole and in this model it is located where
$\partial_+\phi=0$ \cite{rst1}.  The presence of a region of
trapped points can be determined from data on a Cauchy surface
so that, unlike the event horizon, the apparent horizon is defined
locally.

\includefigures\begin{figure}[bt]
\epsfxsize=7.cm \epsfysize=6cm
\epsfbox{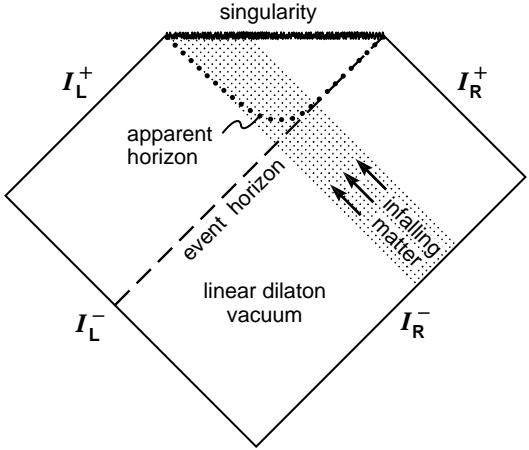}
\caption{Penrose diagram for black hole formed by incoming matter.}
\label{collpenr}
\end{figure}\fi

In the gravitational collapse solution (\ref{collapse}) the
apparent horizon is the curve
\be
x^-_A(x^+)=-P_+(x^+)\,.
\label{apphor}
\ee
In classical solutions the apparent horizon is always spacelike
or null and along it the area of the transverse two-sphere,
${\cal A}_A=\exp{[-2\phi(x^+,x^-_A)]}$,
is a non-decreasing function of $x^+$.  This will no longer hold
when quantum effects are included and the black hole evaporates.

The Penrose diagram in Figure~\ref{collpenr}
depicts black hole formation in
the classical theory.  Both the apparent horizon and the global
event horizon are indicated in the figure.

\section{Semi-Classical Black Hole Physics}
\label{seciii}
Our analysis of the classical CGHS model has revealed
physically interesting geometries including the 1+1-dimensional
counterpart of a black hole formed by the gravitational collapse
of matter.  If a consistent quantum theory of this model can
be constructed it will no doubt be considerably simpler than
a theory of real gravity in 3+1 dimensions, but even the
quantization of two-dimensional gravity is an ambitious goal
and we will have to settle for a few modest steps in that
direction here.

\subsection{The Hawking Effect}
\label{threeone}
In his famous 1975 paper \cite{hawking1} Hawking studied
quantum effects of matter in the classical background
geometry of a black hole formed in collapse and concluded that
the black hole will emit thermal radiation as if it were a
blackbody at a temperature which is proportional to its
surface gravity.  This effect also occurs in 1+1 dimensional
black hole physics \cite{cghs} and for conformally coupled
matter the calculation can be carried out in a neat
fashion by making use of an observation of Christensen and
Fulling \cite{chrful}.  The argument goes as follows.
The classical matter theory of the
$f_i$ fields has a traceless conserved energy-momentum tensor
but when the matter fields are quantized they give rise to a
conformal anomaly,
\be
\label{emttrace}
\langle T_{\>\mu}^{f\>\mu}
\rangle ={N\over 24}R\,.
\ee
The left hand side is the expectation value of the trace of the
matter energy-momentum tensor and on the right hand side
$R$ is the curvature scalar in the classical
background geometry while $N$ is the number of scalar fields
in the theory, or, more generally, the central charge of
the matter system.  In the conformal gauge (\ref{lineel})
the anomaly takes the form,
\be
\langle T^f_{+-} \rangle
= -{N\over 12} \partial_+\partial_-\rho \,.
\label{conftrace}
\ee
If the quantization procedure used for the matter fields is
consistent with general coordinate invariance the energy-momentum
tensor will be conserved, which translates into the following pair
of equations in conformal gauge,
\be
\!\!\begin{array}{l}
0=\partial_-\langle T^f_{++}\rangle +
\partial_+\langle T^f_{-+}\rangle -
2\partial_+\rho\,\langle T^f_{-+}\rangle \,, \\ { } \\
0=\partial_+\langle T^f_{--}\rangle +
\partial_-\langle T^f_{+-}\rangle -
2\partial_-\rho\,\langle T^f_{+-}\rangle \,.
\end{array}
\label{confcons}
\ee
Inserting the conformal gauge expression (\ref{conftrace})
for the anomaly into these equations and integrating gives
\be
\!\!\begin{array}{l}
\langle T^f_{++} \rangle = -{N\over 12}\bigl[
\partial_+\rho\partial_+\rho
- \partial_+^2\rho + t_+(y^+)\bigr] \,, \\ { } \\
\langle T^f_{--} \rangle = -{N\over 12}\bigl[
\partial_-\rho\partial_-\rho
- \partial_-^2\rho + t_-(y^-)\bigr] \,.
\end{array}
\label{fluxes}
\ee
The above argument is quite general as it uses only the
conservation of energy-momentum and the existence of the
conformal anomaly to express all the components of the anomalous
energy-momentum tensor in terms of the background metric.
Some further physical input is needed to fix the functions
of integration $t_\pm$.  In the case of a black hole formed in
collapse this input comes in the form of boundary conditions
imposed at past null infinity,
stating that there is no outgoing energy flux in the initial
vacuum at ${\cal I}^-_L$ and that only the classical matter energy
flux is incident at ${\cal I}^-_R$.

The functions $t_\pm$ are intimately connected with the issue
of regularization in the matter quantum theory.  The
energy-momentum tensor is a composite operator and its expectation
value is not well defined unless we specify a normal ordering
prescription with respect to some vacuum state.  A choice of
vacuum for the matter fields corresponds to a choice of
coordinate system in that the vacuum is defined to contain no
quanta with positive frequency as measured by some time variable.
As a result the normal ordering prescription used to define the
expectation value of the energy-momentum tensor is coordinate
dependent, and observers, which are at rest in a different
reference frame than the one the vacuum is defined in, will
in general measure a non-vanishing energy flux in that vacuum
state.  The role of $t_\pm$ is to keep track of this coordinate
dependent energy flux and their transformation under
a conformal reparametrization is as follows,
\be
t_+(x^+)=\bigl({dx^+\over dy^+}\bigr)^{-2}
\bigl(t_+(y^+)+{1\over 2}{\cal D}^S_{y^+}[x^+]\bigr) \,,
\label{ttransf}
\ee
where
\be
{\cal D}^S_{y}[x]=
{\partial^3 x/\partial y^3\over \partial x/\partial y}
-{3\over 2}
{(\partial^2 x/\partial y^2)^2\over (\partial x/\partial y)^2}
\label{schwartz}
\ee
is the Schwarzian derivative of $x$ with respect to $y$.
A similar relation holds for $t_-$.

We are interested in the outgoing energy flux measured by
asymptotic inertial observers so we want to evaluate $t_\pm$ in a
coordinate system $(\sigma^+,\sigma^-)$ where the metric is
manifestly Minkowskian in the asymptotic regions near
${\cal I}^+_R$ and ${\cal I}^-_R$,
\be
\sigma^+ = \log{x^+} \,,\quad
\sigma^- = -\log{(-x^- - P_\infty)}\,.
\label{tort}
\ee
The boundary conditions on the matter energy-momentum tensor
are applied at past null infinity and imply
$t_\pm(\tilde\sigma^\pm)=0$ in a coordinate system
$(\tilde\sigma^+,\tilde\sigma^-)$ where the metric is manifestly
flat at ${\cal I}^-_L$.  These coordinates are related to the
$\sigma$-coordinates by
\be
\tilde\sigma^+=\sigma^+\,,\quad
e^{-\tilde\sigma^-}=e^{-\sigma^-}+P_\infty \,,
\label{shift}
\ee
and by applying the transformation rule (\ref{ttransf}) one
obtains
\be
\!\!\begin{array}{l}
t_+(\sigma^+)=0 \, \\ { } \\
t_-(\sigma^-)={1\over 4}\bigl((1+P_\infty e^{\sigma^-})^{-2}
-1\bigr) \,.
\end{array}
\label{littlets}
\ee
This in turn implies that there is a non-vanishing outgoing energy
flux at ${\cal I}^+_R$ given by
\be
\langle T^f_{--}\rangle = {N\over 48}\bigl(1-
(1+P_\infty e^{\sigma^-})^{-2}\bigr) \,.
\label{outflux}
\ee
The outgoing energy flux is zero at early retarded times but builds
up to a fixed value and continues forever because our calculation
has not taken into account the back-reaction on the geometry due
to the emitted energy.  The dependence on $P_\infty$ of the energy
flux in (\ref{outflux}) can be removed by a uniform shift of the
retarded time coordinate $\sigma^-$ leaving an expression for the
rate of Hawking radiation which is completely independent of the
original incoming matter energy distribution.

We have seen that two-dimensional black holes emit energy and we
would now like to determine whether the outgoing flux is in the form
of thermal radiation.  One way to
show this is to adapt Hawking's original calculation of Bogolioubov
coefficients to the two-dimensional theory \cite{gidnel} but a
simpler approach utilizes the fact that the matter theory at hand
is a conformal field theory which allows direct calculation of
correlation functions of matter fields in the outgoing radiation.
Consider for example $\langle \partial_-f_i(\sigma^-_1)
\partial_-f_j(\sigma^-_2)\rangle$ at ${\cal I}^+_R$.
Since the matter fields satisfy a free wave equation we can relate
this correlation function to a corresponding one evaluated in the
initial vacuum at ${\cal I}^-_L$, where the $f_i$ are free scalar
fields in Minkowski space,
\beq
\lefteqn{
\langle \partial_-f_i(\sigma^-_1)\partial_-f_j(\sigma^-_2)\rangle
} \nonumber \\
& &=
{\partial\tilde\sigma^-_1\over \partial\sigma^-_1}\,
{\partial\tilde\sigma^-_2\over \partial\sigma^-_2}\,
{\delta_{ij}\over (\tilde\sigma^-_1-\tilde\sigma^-_2)^2} \,.
\label{correlator}
\eeq
In the limit of late retarded time, $\sigma^-\rightarrow\infty$,
this becomes
\be
\langle \partial_-f_i(\sigma^-_1)\partial_-f_j(\sigma^-_2)\rangle
\sim {\delta_{ij}\over
\cosh{(\sigma^-_1{-}\sigma^-_2)}-1} \,,
\label{latetime}
\ee
which is manifestly periodic under the Euclidean time translation
$\Delta\sigma^-\rightarrow\Delta\sigma^- +2\pi i$.
Once the black hole has settled down after its initial formation
it emits thermal radiation at a temperature $T_H=1/2\pi$, which
is independent of the black hole mass.

It is worth noting that although the final answer (\ref{latetime})
is perfectly regular as a function of $\Delta\sigma^-$, as long
as the coordinate difference is not too small, the calculation
nevertheless involves extremely small coordinate differences in the
$\tilde\sigma^-$ frame at an intermediate stage.  This appearance
of extremely short coordinate distances, or equivalently very high
frequencies, is common to many field theoretic calculations of the
Hawking effect \cite{hawking1,jacobson}.

\subsection{The Semiclassical Back-Reaction}
\label{threetwo}
In the previous subsection we have seen energy flux from a black hole
but there was no response in the background geometry.  This was
because we used a classical solution which could not know about the
Hawking effect.  As a remedy for this Callan \etal\ \cite{cghs}
proposed to add to the classical action $S_0$ the Polyakov-Liouville
term, which is induced by quantum effects of the matter,
\beq
\lefteqn{
S_1=-{N\over 96\pi}\int d^2y\sqrt{-g(y)}\int d^2y'\sqrt{-g(y')}
}\nonumber \\
& &  \times R(y)G(y;y')R(y')\,,
\label{liouville}
\eeq
where $G$ is a Green function for the operator $\nabla^2$.
In conformal gauge this non-local term reduces to a local expression,
\be
S_1=-{N\over 12\pi}\int d^2y\,\partial_+\rho\partial_-\rho\,,
\label{s1conf}
\ee
but there remains a residual non-locality in the form of the
functions $t_\pm$, which we encountered before.  Fixing these
functions corresponds to choosing boundary conditions for the Green
function in (\ref{liouville}).

The dilaton gravity sector of the theory also gives rise to quantum
corrections to the effective action but if we take the limit of $N>>24$
the induced term (\ref{liouville}) will dominate over
other one-loop corrections and we need not be concerned with a number
of thorny issues involving functional measures and reparametrization
ghosts.  Those problems will be addressed in Section~\ref{seciv} but
for now we will work in the large $N$ limit.

The semiclassical CGHS equations, obtained by varying
the effective action $S=S_0 + S_1$, have been analyzed by a number of
authors \cite{cghs,bddo,rst1,bghs,lslt1,hawking4}.  They have solutions
describing evaporating black holes and the back-reaction
on the geometry is expected to be reliably described for most
of the lifetime of black holes formed with mass $M>>N$.
The CGHS model cannot be solved exactly at the
semiclassical level and only permits analytical study of the onset of
the evaporation process.  The equations have been solved numerically
to follow the evolution of the geometry \cite{numerical}.

\subsection{The RST Model}
\label{threethree}
Fortunately the semi-classical theory can be modified in such a way
that explicit analytic solutions which exhibit black hole evaporation
are obtained, as was first shown by Bilal and Callan
\cite{bilcal} and de~Alwis \cite{dealwis}.  A particularly simple
semiclassical model of this type was introduced by Russo, Susskind,
and Thorlacius (RST) \cite{rst2}, who proposed to include in the
effective action, in addition to the non-local term (\ref{liouville}),
the local term
\be
S_2 = {N\over 96\pi}\int d^2y\,\sqrt{-g}\,R\phi\,,
\label{rstterm}
\ee
which takes the form
\be
S_2 = -{N\over 12\pi}\int d^2y\,\phi\partial_+\partial_-\rho\,.
\label{confrst}
\ee
in conformal gauge.  The role of this
term is to restore at the semiclassical level the symmetry, generated
by the conserved current $j^\mu=\partial^\mu(\rho-\phi)$, which
enabled the exact solution of the classical theory.  The new
term in the action is manifestly covariant so it is allowed by the
symmetries of the original theory and could have been included from
the beginning.  It has the appearance of a one-loop counterterm and
therefore it does not disturb the classical physics of the model in
the asymptotic region where $e^{-2\phi}>>N/24$.

The analysis of the model is simplified if we
introduce new field variables for the dilaton gravity
sector,\footnote{We adapt here the conventions of \cite{strominger1}
which are well suited to taking the large $N$ limit.}
\be
\!\!\begin{array}{l}
\Omega={12\over N}e^{-2\phi}+{1\over 2}\phi
-{1\over 4}\log{N\over 48} \,, \\ { } \\
\chi=\rho + {12\over N}e^{-2\phi}-{1\over 2}\phi
+{1\over 4}\log{N\over 3} \,,
\end{array}
\label{newvar}
\ee
for which the effective action $S=S_0+S_1+S_2$ takes the form
\beq
\lefteqn{
S={N\over 12\pi}\int d^2y \bigl[
-\partial_+\chi\partial_-\chi
+\partial_+\Omega\partial_-\Omega
}\nonumber \\
& &
+e^{2(\chi-\Omega)}+{6\over N}\sum_{i=1}^N
\partial_+f_i\partial_-f_i
\bigr]  \,.
\label{effact}
\eeq
In terms of the new variables the equations of motion are
\be
\!\!\begin{array}{l}
\partial_+\partial_-f_i =0 \,\\ { } \\
\partial_+\partial_-\chi=\partial_+\partial_-\Omega
=-e^{2(\chi-\Omega)} \,,
\end{array}
\label{rsteqs}
\ee
and the semiclassical constraint equations become
\be
(\partial_\pm\chi)^2-\partial_\pm^2\chi
-(\partial_\pm\Omega)^2 + t_\pm = \tilde T_{\pm\pm} \,.
\label{rstconstr}
\ee
Here $\tilde T_{\pm\pm}$ is the observable energy-momentum flux
in the asymptotic region, rescaled by a factor of $12/N$,
which is appropriate in the large $N$ limit where we study black
holes formed by incoming energy measured in units of $N$.

We can choose Kruskal coordinates, in which $\chi=\Omega$, and
the equations of motion and constraints reduce to
\be
\!\!\begin{array}{l}
\partial_+\partial_-\Omega=-1 \,, \\ { } \\
-\partial_\pm^2\Omega=\tilde T_{\pm\pm}-t_\pm \,.
\end{array}
\label{kruskeqs}
\ee
Notice the similarity with the classical equations
(\ref{kruskeqs}).
The matter energy-momentum tensor $\tilde T_{\pm\pm}$ is normal
ordered with respect to the vacuum state appropriate to inertial
observers in the asymptotically Minkowskian coordinates
(\ref{tort}).  The vacuum state has $\tilde T_{\pm\pm}=0$ and
$t_\pm(\sigma^\pm)=0$, which gets transformed to
$t_\pm(x^\pm)=1/4{x^\pm}^2$ in Kruskal coordinates under
(\ref{ttransf}).  The vacuum solution obtained by
integrating (\ref{kruskeqs}) is then
\be
\Omega=-x^+x^- - {1\over 4}\log{(-x^+x^-)}
-{1\over 2}\log{2} \,.
\label{lindil}
\ee
A comparison with the field redefinition (\ref{newvar}) reveals
that in this semiclassical model the dilaton field is linear,
$\phi=-\sigma^1-{1\over 2}\log{(N/12)}$, in the vacuum solution
just as in the classical theory.

\begin{figure}[bt]
\epsfxsize=5.cm \epsfysize=5cm
\epsfbox{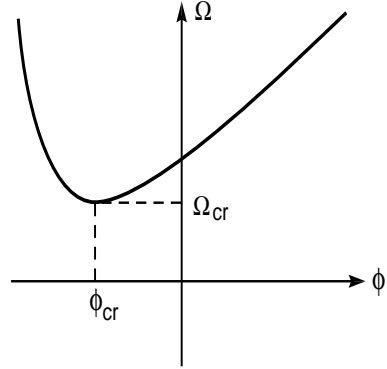}
\caption{The field redefinition from $\phi$ to $\Omega$ is not
one-to-one.}
\label{redef}
\end{figure}

Now consider a geometry with leftmoving matter incident on the
vacuum from ${\cal I}^-_R$.  The semiclassical solution is
\beq
\lefteqn{
\Omega=-x^+\bigl(x^-+\tilde P_+(x^+)\bigr) + \tilde M(x^+)
} \nonumber \\
& &
-{1\over 4}\log{(-x^+x^-)} -{1\over 2}\log{2} \,,
\label{nobc}
\eeq
where $\tilde M$ and $\tilde P_+$ are the moments (\ref{energy})
and (\ref{kruskalmomentum}) of the incoming energy flux in
Kruskal coordinates, rescaled by a factor of $12/N$.
Although this is a perfectly good solution of the semiclassical
equations (\ref{kruskeqs}) its physical interpretation is
problematic.  The reason is that the range of values taken by
$\Omega$ as a function of $x^+$ and $x^-$ is unrestricted but
the field redefinition (\ref{newvar}) is degenerate (see
Figure~\ref{redef}), and $\Omega$ below a certain critical
value $\omcr$, corresponds to a complex value of the original
dilaton field. The critical point, where $\Omega'(\phi)=0$, is
at $\phicr = -{1\over 2}\log{(N/48)}$ and $\omcr ={1\over 4}$.

\includefigures\fi

The existence of this critical value of the dilaton field has
important implications.  By using the equations of motion
(\ref{rsteqs}) written in terms of $\rho$ and $\phi$ one can
express the spacetime curvature as
\be
R={4\over 1-{N\over 48}e^{2\phi}}
\bigl(1-(\nabla\phi)^2\bigr)\,,
\label{curvature}
\ee
and we see that in general the curvature will diverge where
$\phi=\phicr$, even if the solution for $\chi$ and $\Omega$ is
perfectly regular there.  Gravitational quantum corrections are
strong in this model when the dilaton field is near its critical
value.  This can be seen by defining the two-component vector
\be
\Phi=\left[\begin{array}{c}\phi \\ \rho\end{array}\right]\,,
\label{twocomp}
\ee
and assembling the kinetic terms in the full effective action
$S=S_0+S_1+S_2$ into $(\partial_+\Phi)M(\partial_-\Phi)$.
The role of gravitational coupling is played by
\be
(-\det{M})^{-1/4}\sim (e^{-2\phi}-{N\over 48})^{-1/2},
\label{coupling}
\ee
which goes to infinity as $\phi\rightarrow\phicr$.

It is tempting to ignore the problem of unphysical values of
$\Omega$ and simply define the semiclassical theory in terms of
the effective action (\ref{effact}) for $\chi$ and $\Omega$.
Such a theory has the appropriate classical limit by construction
and the smooth behavior of $\chi$ and $\Omega$ in the strong
coupling region of the original theory in effect resolves
the classical singularity.  Unfortunately this approach is
undermined by an instability.  The incoming
matter excites the system from its vacuum configuration and
Hawking radiation is emitted to ${\cal I}^+_R$.  The mismatch
between inertial coordinates at ${\cal I}^+_R$ and ${\cal I}^-_L$
is given by (\ref{shift}), with $P_\infty$ replaced by
$\tilde P_\infty$,
and the calculation of the Hawking flux at ${\cal I}^+_R$
proceeds in the same manner.  Although the semiclassical
solution exhibits a back-reaction effect on the geometry due to
the Hawking emission there is nothing to
turn the outgoing flux off when the emitted energy
exceeds the total incoming energy and the Bondi mass measured at
${\cal I}^+_R$ goes to negative infinity at late times.

In order to avoid these problems of unphysical $\Omega$ values
and negative energy instability, Russo \etal\ \cite{rst2}
interpreted the curve $\Omega=\omcr$ as the analog of the origin
of radial coordinates in higher dimensional gravity, beyond
which solutions should not be continued, and proposed
`phenomenological' boundary conditions for $\Omega$,
\be
\partial_+\Omega\big\vert_{\Omega=\omcr}=0=
\partial_-\Omega\big\vert_{\Omega=\omcr} \,,
\label{rstbc}
\ee
which ensure that the spacetime curvature remains finite at
the critical curve where it is timelike.  This turns out to
stabilize the semiclassical evolution, which is perhaps not
surprising since negative energy configurations typically have
naked singularities and the above boundary conditions implement
a form of cosmic censorship in the two-dimensional theory.

It should be noted, however, that these boundary conditions for
$\Omega$ are not the most general ones allowed and they do not
imply boundary conditions for the matter fields, which is a
drawback if we want to discuss the quantum state of
the outgoing matter in connection with the information paradox.
It was initially claimed \cite{rst2} that the RST boundary
conditions on $\Omega$ would be compatible with Dirichlet or
Neumann boundary conditions on the $f_i$ but it was later
realized that this is not the case \cite{strtri,stu}, and
(\ref{rstbc}) may in fact not be realizable as the
semiclassical limit of any consistent quantum mechanical
boundary conditions.  In Section~\ref{seciv} we will discuss
alternative choices of boundary conditions
\cite{chuver,dasmuk,aslt}, which are compatible with simple
reflecting conditions on the matter fields, but these models
are somewhat more complicated than the RST model and the
analysis of the semiclassical solutions less transparent.
We will therefore explore the physical picture presented in the
RST model before moving on to other models.

\subsection{Semiclassical Black Holes}
\label{threefour}
Let us first consider the static solutions of the semiclassical
equations (\ref{kruskeqs}) subject to the boundary conditions
(\ref{rstbc}),
\beq
\lefteqn{
\Omega=-x^+x^- - {(1-a)\over 4}\log{(-x^+x^-)}
}\nonumber \\
& &
+ M + {a\over 4}
+ {(1-a)\over 4} \log{({1-a\over 4})} \,.
\label{statbh}
\eeq
These static geometries are characterized by two parameters.
One is proportional to the asymptotic energy density,
$a/4=\tilde T_{++}=\tilde T_{--}=$ as
$\sigma^1\rightarrow\infty$, and the other one $M$,
will be referred to as the mass even if the
canonical ADM mass diverges for geometries with a non-vanishing
asymptotic energy density.\footnote{One might expect a disastrous
back-reaction on the geometry in the asymptotic region,
corresponding to the Jeans instability in 3+1-dimensional gravity,
but this is avoided because the coupling strength $e^\phi$ goes
to zero there.}

The solution with $a=0$ and $M=0$ is the linear dilaton
vacuum (\ref{lindil}), for $a=0$ and $M>0$ it is
a `quantum kink' solution with a singular horizon at
$x^+x^-=0$ \cite{bghs,lslt1}, and for $a=0$ and
$M<0$ it has a naked singularity.

A solution with $0<a<1$ and $M=0$ corresponds to a
heat bath at a temperature $T=a/2\pi$.
A semiclassical black hole emits Hawking radiation and a static
configuration can only exist if the black hole is in equilibrium
with a heat bath at a temperature equal to the Hawking temperature
$T_H=1/2\pi$.  This is described by a static solution with
$a=1$ and $M>0$, which has a spacelike singularity at
$x^+x^-=M$ and non-singular event horizon at $x^+x^-=0$.
Notice that since the boundary curve $\Omega=1/4$ is spacelike in
the static black hole geometries they are
determined without applying the boundary conditions (\ref{rstbc}).

The black hole temperature is independent of the mass parameter
so that the specific heat is infinite.  Random fluctuations in the
thermal flux of energy at the horizon will therefore cause the black
hole mass to slowly increase or decrease with time \cite{stu,ssstt}.
In ordinary systems, which have a finite positive specific heat,
such fluctuations are stabilized by a response in the temperature
of the system.  A momentary increase (decrease) in the energy of a
system in equilibrium with a heat bath causes an increase (decrease)
in the temperature of the system, which in turn causes heat to flow
to (from) the bath.  In the case at hand, the black hole
temperature does not respond to the energy fluctuation and there
is no restoring effect to maintain a balance.  With time the
black hole mass will therefore random walk away from its
original value.  The semiclassical equations do not incorporate
this thermal effect but physically the one-parameter family of
distinct static black hole solutions should be replaced by a
single ensemble which includes black holes of arbitrary mass.

\begin{figure}[bt]
\epsfxsize=7.5cm \epsfysize=6cm
\epsfbox{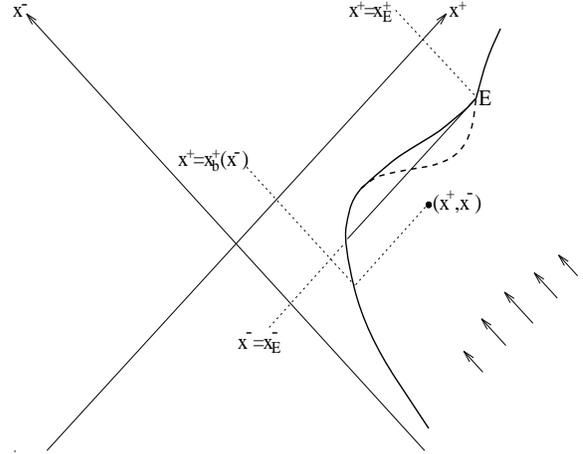}
\caption{Kruskal diagram for a black hole formed by
incoming matter in the semiclassical theory.}
\label{bhform}
\end{figure}

We now turn our attention to dynamical solutions of the
semiclassical equations subject to the boundary conditions
(\ref{rstbc}).  The semiclassical geometry can be explicitly
determined everywhere in spacetime and expressed in a
relatively compact form,
\beq
\lefteqn{
\Omega=-x^+\bigl(x^-+\tilde P_+(x^+)\bigr) + \tilde M(x^+)
} \nonumber \\
& &
-\tilde M\bigl(x^+_b(x^-)\bigr)
-{1\over 4}\log{\bigl(x^+/x^+_b(x^-)\bigr)}\,.
\label{gensol}
\eeq
Here $x^+_b(x^-)$ is the $x^+$ value of the point on the
boundary curve $\Omega=\omcr$ from which the reflected signal
propagates to $(x^+,x^-)$ as shown in Figure~\ref{bhform}.

\includefigures\fi

The qualitative behavior of the semiclassical solution
(\ref{gensol}) depends on the incoming matter energy.  There
is a threshold energy flux required for black hole formation,
which coincides with the rate of Hawking emission from a
black hole.  As long as the incoming energy flux remains
below threshold, $\tilde T_{++}<1/4{x^+}^2$, the boundary
curve, defined by $\Omega(x^+_b,x^-_b)=1/4$, will be timelike
and its shape given by
\be
x^-_b = -\tilde P_+(x^+_b) -{1\over 4x^+_b} \,.
\label{boundary}
\ee

Consider a geometry where the incoming energy flux tapers off
at early and late times and always remains below threshold.
As $x^+\rightarrow\infty$ and $x^-\rightarrow 0$ the solution
(\ref{gensol}) approaches the linear dilaton vacuum
(\ref{lindil}) up to a uniform shift of $x^-$ by
$\tilde P_\infty$.  By combining the boundary conditions
(\ref{rstbc}) and the constraint equations in (\ref{kruskeqs})
one can obtain the outgoing energy flux due to a given
incoming matter energy profile.  The total outgoing energy
can then be computed by integrating the outgoing flux in the
asymptotically inertial coordinate system (\ref{tort}) over
all retarded time (remembering to take into account the
anomalous transformation properties of $t_-$ when passing
from the Kruskal coordinate system) and one finds that it
equals $\tilde M_\infty$, the total incoming energy.  The RST
boundary conditions thus respect overall energy conservation
even if it is not manifest.

\begin{figure}[bt]
\epsfxsize=7cm \epsfysize=7cm
\epsfbox{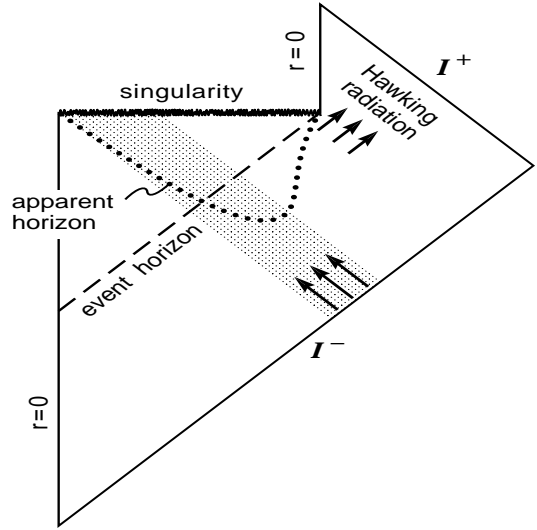}
\caption{Penrose diagram for a black hole formed by
incoming matter in the semiclassical theory.}
\label{scpenr}
\end{figure}

Now consider the case when the incoming energy flux becomes
larger than the threshold value at some point.  Then the
boundary curve becomes spacelike and boundary conditions
can no longer be applied there.  A spacelike segment of the
boundary is a curvature singularity.  It forms
inside a region of future trapped points which is bounded on
the outside by an apparent horizon, located where
$\partial_+\phi=0$, as shown in Figures~\ref{bhform}
and~\ref{scpenr}.

The apparent horizon curve $(x^+_a,x^-_a)$ satisfies
\be
x^-_a = -\tilde P_+(x^+_a) -{1\over 4x^+_a} \,,
\label{apho}
\ee
which is the same equation as (\ref{boundary}), which
determined the timelike boundary, but the two apply under
different circumstances as the apparent horizon only
exists for those values of $x^+$ where the boundary curve is
spacelike and (\ref{boundary}) does not hold.
The apparent horizon curve is itself spacelike whenever the
incoming energy flux is above threshold and the black
hole is gaining mass.  Once the incoming flux falls below
threshold there is a net loss of energy from the black hole
due to Hawking emission and the apparent horizon becomes
timelike.  If the incoming flux remains below threshold for
a sufficiently long time the apparent horizon will run
into the spacelike singularity.  At the
black hole endpoint, which is denoted by $E$ in
Figure~\ref{bhform}, the boundary curve becomes timelike again
and the boundary conditions (\ref{rstbc}) can be applied.

The solution in the causal future of the black hole
endpoint matches continuously onto the evaporating black
hole solution across the line segment $\{x^-=x^-_E,
x^+>x^+_E\}$, but the match is not smooth.  The derivative
$\partial_-\Omega$ is discontinuous there by an amount which
by the constraint equations in (\ref{kruskeqs})
corresponds to an outgoing shock wave carrying a small
negative energy.  The fact that negative energy is carried
out from the endpoint is perhaps strange but it is not very
serious.  Energy density is not positive definite in
quantum theories and global energy positivity is not violated
by this negative energy `thunderpop' whose energy is bounded
by the analog of the Planck scale in this theory,
$\vert \tilde T_{--}(\sigma^-)\vert <1/4$.

The expression for spacetime curvature (\ref{curvature})
takes a simple form on the apparent horizon curve
\be
R={4\over 1-{N\over 48}e^{2\phi}} \,,
\label{horcurv}
\ee
and it follows that the curvature diverges on the apparent horizon
as it approaches the black hole singularity.  The apparent horizon
of an evaporating black hole is visible from future null infinity
and the diverging curvature means that the endpoint of evaporation
is a naked singularity.  Cosmic censorship is therefore violated
but by adopting the boundary conditions (\ref{rstbc}) after the
evaporation is complete the violation is kept to a minimum.

\includefigures\fi

Consider, for concreteness, a geometry where a black hole is formed
by an incoming energy flux which is above threshold for a while but
is then turned off at some finite value of $x^+$.
The associated Penrose diagram is shown in Figure~\ref{scpenr}.
In this case, the
black hole endpoint occurs in a region where there is no incoming
energy flux and once the thunderpop has been emitted the solution
(\ref{gensol}) reduces to the linear dilaton vacuum (\ref{lindil})
up to the usual shift of $x^-$ by $\tilde P_\infty$.  The outgoing
energy flux in Hawking radiation takes exactly the same form as in
(\ref{outflux}) except that it shuts off once the thunderpop
arrives at ${\cal I}^+_R$.  In a classical background geometry
the rate of Hawking emission did not depend at all on the
incoming energy profile.  We are doing somewhat better here in
that the radiation only lasts for a time which is proportional to
the original black hole mass, so that energy is conserved, but
apart from that the Hawking energy flux reflects none of the
detailed structure of the incoming matter distribution.

Since we do not have boundary conditions on the matter fields we
cannot make precise statements about information loss at this
stage.  On the other hand, since all the incoming matter that
arrives between $x^+=x^+_b(x^-_E)$ and $x^+=x^+_E$ will pass
through the global event horizon of the black hole without any
chance of reflecting off the boundary, it seems clear that no
local boundary conditions on the $f_i$ can prevent information
from entering the black hole in this model.

We would be able to make a stronger statement if we could impose
simple reflecting boundary conditions, either Dirichlet or
Neumann, on the matter fields and appeal to the standard lore
\cite{mirrors} on quantum fields reflecting off moving
mirrors.  The problem is that the RST boundary conditions
(\ref{rstbc}) are incompatible with reflecting boundary conditions
for the matter fields \cite{strtri,stu} and it is therefore in
general inconsistent to use the above semiclassical solutions as
background geometries for quantized matter fields.

The incompatibility can be seen by comparing the outgoing energy
flux found for a semiclassical solution (\ref{gensol}) in the
low-energy sector of the RST model to the energy flux that would be
obtained if the timelike boundary curve were replaced by a mirror
that moved along the same trajectory.  It is straightforward to
show that the two disagree and that the disagreement becomes
particularly pronounced in the limit when the incoming matter
energy flux approaches the threshold for black hole formation
\cite{stu}.
If we go beyond the the threshold and let a black hole form then
the problem gets even worse.  In this case the reflecting boundary
is disconnected as it jumps from $(x^+_b,x^-_E)$ to the black hole
endpoint at $(x^+_E,x^-_E)$.  A disconnected mirror trajectory
gives rise to an infinite burst of outgoing energy at the
discontinuity \cite{anddew} instead of the relatively benign
thunderpop found in the RST model.

This incompatibility is a stumbling block that confounds
the quantization of the RST model.  Alternate boundary conditions
for dilaton gravity, which are consistent with Dirichlet or
Neumann conditions on the matter fields, have been considered
\cite{chuver,dasmuk,aslt}, and will be discussed below.

\section{Conformally Invariant Models}
\label{seciv}

So far, we have considered quantum effects of the matter
theory but treated the dilaton and conformal factor as classical
fields whose equations of motion receive correction terms due
to the quantization of matter fields.  While the contribution
from the matter to the effective action dominates over other
one-loop term in the large $N$ limit we must nevertheless
consider a more systematic quantization of the complete theory
if we wish to address fundamental questions such as information
loss.  Another reason to go beyond the large $N$ approximation
is that it breaks down near the singularity inside a black hole
and at the endpoint of black hole evaporation.
These issues have been addressed by a number of authors
\cite{bilcal,dealwis,jrat,gidstr2,burcha}.  We will adopt the
conventions of \cite{gidstr2} in our discussion.

\subsection{Conformal Gauge Quantization}
\label{fourone}
The quantum theory of dilaton gravity coupled to scalar fields
is formally defined in terms of the functional integral
\be
{\cal Z} = \int {\cal D}(g,\phi,f_i)\,
e^{iS_0} \,.
\label{pathint}
\ee
In order to be meaningful this expression requires a prescription
for gauge fixing, regularization, and renormalization.  It is
usually assumed that some covariant non-perturbative method of
regularizing the continuum theory exists although at present no
such method is available.  Instead we shall rely on a procedure
which is analogous to old-fashioned methods of regularization
and renormalization in gauge theories. By this procedure
the theory is first regularized in a non-covariant way and then
one compensates for the resulting non-invariance by allowing
the effective Lagrangian to contain terms that are not gauge
invariant.  At the end of the day the gauge
symmetry is re-imposed through Ward identities, which constrain
the added terms.  A version of this method applied to
two-dimensional gravity is as follows \cite{cgquant}.

The first step is gauge fixing.  We have to remove the
over-counting of metrics due to general coordinate invariance.
This is achieved by fixing some reference metric $\hat g_{ab}$,
which could for example be the flat Minkowski metric, and
then choosing coordinates such that the physical metric is
conformal to the reference metric,
\be
g_{ab}=e^{2\rho}\hat g_{ab} \,.
\label{refmet}
\ee
In two spacetime dimensions such a coordinate system can always
be found locally.  The original path integral over metrics is
replaced by an integral over the conformal factor $\rho$ along
with the usual anticommuting Faddeev-Popov ghosts.

The next step is regularization.  The theory has ultraviolet
divergences which must be regularized in order to define the
gauge fixed path integral.  A non-perturbative regulator could
for example be introduced by discretizing the spacetime so that
there is a shortest length, as measured by the reference metric,
\be
\hat g_{ab}\,\delta x^a\,\delta x^b > \varepsilon \,,
\label{cutoff}
\ee
where $\delta x^a$ is the element connecting nearest neighbor
lattice points and $\varepsilon$ tends to zero as the cutoff is
removed.  A more covariant method would refer the cutoff to the
physical metric $g_{ab}$, but then the regularization would
depend on the conformal factor, which is one of the fields being
integrated over, and the regularized path integral would not
have a concrete definition.

The final step involves renormalization.  Performing the path
integral over short distance fluctuations of the conformal
factor, dilaton, and matter fields generates various interaction
terms, involving $\rho$, $\phi$, and $f_i$, in the effective
Lagrangian.  In general these terms depend on the arbitrarily
chosen reference metric $\hat g_{ab}$ and therefore the
effective action will not be manifestly covariant.  On the
other hand, the original theory is assumed to be invariant
under general coordinate transformations so we must impose on
the renormalized theory that the value of the path integral
does not depend on the choice of reference metric.
This can be achieved by, first of all, arranging the terms in
the effective action to be covariant with respect to
$\hat g_{ab}$.  This does not restrict the possible couplings
but merely labels them by their transformation properties
under reparametrizations of the reference system.  Then we
impose that the path integral remains invariant under the
transformation
\be
\hat g_{ab}\rightarrow e^{2\delta\alpha}\hat g_{ab}\,\quad
\rho\rightarrow \rho-\delta\alpha\,,
\label{fidtrans}
\ee
which leaves the left hand side of (\ref{refmet}) invariant.
This condition translates into the restriction that the
beta-functions of all the couplings in the
theory must vanish.  In other words, the gauge fixed theory
must be an exact fixed point of the renormalization group in
order to maintain the original general covariance.  Since
the Faddeev-Popov ghosts contribute $-26$ to the conformal
anomaly the theory of the remaining fields should be a
$c=26$ conformal field theory.

We are thus led to consider a reparametrization
invariant field theory in two dimensions, where the action
can {\it a priori\/} include terms with an arbitrary functional
dependence on the $N+2$ fields $X^\mu =(\rho,\phi,f_i)$, and
with any number of derivatives acting on the fields,
\beq
\lefteqn{
S=-{1\over 2\pi}\int d^y\sqrt{-\hat g}\bigl[
T(X) +{1\over 2}\hat R\,\Phi(X)
}\nonumber \\
& &
+\hat g^{ab}\hat\nabla_aX^\mu\hat\nabla_bX^\nu G_{\mu\nu}(X)
+\ldots\bigr] \,.
\label{genact}
\eeq
We have only written terms of scaling dimension zero and two
but in general there is an infinite sequence of possible
couplings involving any number of derivatives of the $X^\mu$
and higher powers of $\hat R$, the curvature of the reference
metric.

This class of theories has been extensively studied in string
theory where the action (\ref{genact}) describes strings in
background fields in $N+2$ dimensional target space.  The
beta-function equations, which implement conformal invariance
of the two-dimensional theory, have the form of field
equations in target space.  In order to completely specify
the two-dimensional theory we need to give initial data for
all the beta-function equations and then solve them.  Exact
solutions are only available under very special circumstances
and in the general case we can at best hope to find a small
parameter for a perturbation expansion.  In the case at hand,
we have the loop expansion parameter $e^{2\phi}$ of the dilaton
gravity which is small in the classical limit.

\subsection{Models for Black Hole Physics}
\label{fourtwo}
The conformal invariance condition places restrictions on the
possible two-dimensional effective theories that correspond to
generally covariant theories of gravity, but there is still an
infinite class of allowed theories.  The different theories
correspond to different physical systems and we can narrow the
field down by imposing some further physical restrictions,
which are appropriate to the application we have in mind.

First of all, since we are interested in studying black holes
we require our effective theory to reduce to the classical
CGHS theory in the limit $e^{2\phi}\rightarrow 0$.
As a corollary to this requirement we also want the linear
dilaton vacuum, or a close relative, to be a solution of the
effective theory so that we can study gravitational collapse as
in the CGHS model.

As a further requirement we want the leading order corrections
in powers of $e^{2\phi}$ to give rise to the appropriate
semiclassical behavior of black holes, \ie\ Hawking radiation
and its back-reaction on the geometry.  A subtle issue, which
has led to persistent confusion in the literature, arises in
this context.  The Hawking radiation should consist only of
matter fields, whereas the non-propagating fields of the theory,
$\rho$, $\phi$, and the Faddeev-Popov ghosts, should not
contribute to the outgoing energy flux.  We have already
encountered the Polyakov-Liouville term (\ref{liouville}),
which is responsible for the matter contribution to the Hawking
effect, and naively one would expect analogous terms arising as
one-loop contributions from the path integral over $\rho$,
$\phi$, and the ghosts.  The coefficient in front of the
Polyakov-Liouville term in the effective action would then
be proportional to $N+2-26$ instead of $N$ and this would lead
to the unphysical conclusion that the rate of Hawking evaporation
is not proportional to the number of available channels $N$ but
rather $N-24$.  This can be ignored in the large $N$
limit, as we did in Section~\ref{seciii}, but for finite $N$,
and $N<24$ in particular, this is not acceptable.

This problem was solved at the one-loop level by Strominger
\cite{strominger2} who introduced local covariant counterterms
into the effective action to decouple the non-propagating
modes from the Hawking radiation.  His choice of counterterms
was motivated by the observation that the natural metric that
defines the functional measure for $\rho$, $\phi$, and the
ghost fields in the path integral involves $\rho-\phi$ rather
than just $\rho$.  This can be seen by
comparing the kinetic terms of the dilaton gravity part of
the classical action (\ref{classact}) to the matter kinetic
terms.  The resulting one-loop contribution to the effective
action due to $\rho$, $\phi$, and the ghosts takes the
form
\be
S_{gh}={24\over 12\pi}\int d^2y\,
\partial_+(\rho-\phi)\partial_-(\rho-\phi)
\label{ghostpl}
\ee
in conformal gauge.  The $\partial_+\rho\partial_-\rho$ part is
non-local in a general gauge and combines with the matter
contribution to the
Polyakov-Liouville counterterm of the model but the other terms
in (\ref{ghostpl}) correspond to local counterterms involving
$R\phi$ and $(\nabla\phi)^2$.  The ultimate justification for
this choice comes when we consider semiclassical black holes in
our effective theory and determine their rate of Hawking
evaporation.

To summarize, we will restrict the form of our effective action
to be such that it defines a $c=26$ conformal field theory and
in the limit $e^{2\phi}\rightarrow 0$ the leading order terms
reduce to the classical CGHS model (\ref{classact}), corrected
by (\ref{s1conf}) and (\ref{ghostpl}).\footnote{We do not include
the RST term (\ref{confrst}), which served as a means to obtain
exactly soluble semiclassical equations in Section~\ref{seciii},
but would not simplify the analysis here.}
This requires the target space fields that appear in
(\ref{genact}) to have the following limiting behavior
\cite{gidstr2}, up to $O(e^{2\phi})$ terms,
\be
\!\!\begin{array}{ll}
G_{\rho\rho}\sim -\gamma\,, &
G_{\rho\phi}\sim 2e^{-2\phi}{-}2 , \\ { } & { } \\
G_{\phi\phi}\sim -4e^{-2\phi}+2\,, &
G_{ii}\sim {1\over 2}\,,  \\ { } & { } \\
\Phi\sim -2e^{-2\phi}{-}4\phi{-}2\gamma\rho , &
{ }  \\ { } & { } \\
T\sim -4e^{2\rho-2\phi}+O(1)\,, & { }
\end{array}
\label{targetfields}
\ee
where we have defined $\gamma=(N-24)/12$.

As a final requirement on our effective action we would like
to impose the `theoretician's condition' that it be possible to
explicitly solve the semiclassical equations of the model.
This is clearly not a physical requirement but if it can be met
it allows detailed analytical analysis of the semiclassical
physics, which enhances our understanding of the model, and
further down the road it may also simplify some technical steps
involved in its quantization.

\subsection{A Soluble Model}
\label{fourthree}
It turns out to be possible to satisfy all the above requirements
and construct a conformally invariant model, where the semiclassical
equations can be solved exactly, as was first shown independently
by Bilal and Callan \cite{bilcal} and de~Alwis \cite{dealwis}.
Consider the effective action\footnote{The formalism we are using
handles $\gamma>0$ and $\gamma<0$ simultaneously.  The special
case $\gamma=0$ requires separate treatment and has been
considered by a number of authors.
See \eg\ \cite{dasmuk,verl}.}
\beq
\lefteqn{
S={1\over \pi}\int d^2y \bigl[
-\gamma\partial_+X\partial_-X + \gamma\partial_+Y\partial_-Y
} \nonumber \\
& &
+ e^{2(X-Y)} + {1\over 2}\sum_{i=1}^N \partial_+f_i\partial_-f_i
\bigr] \,,
\label{xyact}
\eeq
where the field variables $X$ and $Y$ are related to the conformal
factor and dilaton as follows,
\beq
\lefteqn{
\!\!\begin{array}{l}
X=\rho+{1\over \gamma}(e^{-2\phi}+2\phi) \,, \\ { } \\
Y=-{1\over \gamma}\int d\phi \sqrt{
4e^{-4\phi}-4(\gamma{+}2)e^{-2\phi}+2(\gamma{+}2)} \,.
\end{array}
} \nonumber \\
& & { }
\label{xydef}
\eeq
Substituting $X$ and $Y$ into this action leads to the classical
CGHS action (\ref{classact}) corrected by the one-loop terms
(\ref{s1conf}) and (\ref{ghostpl}) along with additional potential
terms which are subleading in the loop expansion parameter
$e^{2\phi}$.  The $X,Y$ theory is identical to the $\chi,\Omega$
theory of the RST model (up to some numerical factors) but the
field redefinition to $\rho$ and $\phi$ variables differs
significantly and this has important consequences for the physics.

The gravitational part of the
energy-mo\-men\-tum tensor is given by
\be
T^g_{\pm\pm}=\gamma (\partial_\pm Y\partial_\pm Y
-\partial_\pm X\partial_\pm X +\partial^2_\pm X) \,,
\label{gravstress}
\ee
and together with the matter part it generates two independent
$c=26$ Virasoro algebras.

The semiclassical equations of motion of this theory are obtained
by varying the action (\ref{xyact}) and up to factors of $\gamma$
they are identical to the equations of motion (\ref{rsteqs}) of
the RST model.  As before, we can work in Kruskal gauge,
$X=Y$, and express the general static solution in terms of two
parameters,
\be
Y=-{1\over \gamma}x^+x^- +(p-{1\over 4})\log{(-x^+x^-)}
+{\mu\over \gamma} \,.
\label{ystat}
\ee
The vacuum configuration has $\mu=0$ and $p=-1/2\gamma$
\cite{gidstr2}.  In order to see why this choice of $p$ is natural
it is instructive to evaluate the gravitational part of the
energy-momentum tensor for the vacuum configuration in
asymptotically Minkowskian coordinates, which are related to
the Kruskal coordinates by $s^\pm=\pm\log{x^\pm}$.  The result is
\be
T^g_{++}(s^+)={1\over 2} \,,
\label{stressflux}
\ee
along with a similar expression for $T^g_{--}$.  The non-vanishing
right hand side cancels agains a ghost energy flux,
$T^{gh}_{++}=2\bigl(\partial_+(\rho-\phi)\bigr)^2$, which is
included in $T^g_{++}$ on the left hand side of (\ref{stressflux}).

This choice of $p$ is also the only one which leads to the correct
rate of Hawking evaporation of a black hole formed by collapse of
matter into the vacuum \cite{aslt}.  To see that, keep the parameter
$p$ arbitrary for the time being and consider a dynamical solution
with leftmoving matter incident on the vacuum from ${\cal I}^-_R$,
\beq
\lefteqn{
Y=-{1\over \gamma}x^+\bigl(x^-+P_+(x^+)\bigr)
+ {1\over \gamma}M(x^+)
} \nonumber \\
& &
+ (p-{1\over 4})\log{(-x^+x^-)} \,.
\label{ydyn}
\eeq

The evaporation rate can be obtained by transforming to the
inertial coordinate system (\ref{tort}) at ${\cal I}^+_R$ and
evaluating the Bondi mass \cite{cghs},
\be
m(\sigma^-)=2e^{\sigma^+-\sigma^-}
(\delta\rho+\partial_+\delta\phi-\partial_-\delta\phi) \,.
\label{bondi}
\ee
Here $\delta\rho$ and $\delta\phi$ are the deviations of $\rho$
and $\phi$ from their vacuum values in (\ref{ydyn}).  After some
algebra one finds
\beq
\lefteqn{
m(\sigma^-)=M_\infty
+(p-{1\over 4})\gamma \bigl[
\log{(1{+}P_\infty e^{\sigma^-})}
} \nonumber \\
& & \qquad
+ {P_\infty \over P_\infty +e^{\sigma^-}}\bigr].
\label{bondi2}
\eeq
If we choose $p=-1/2\gamma$ then the Bondi mass decays at a rate
\be
{dm\over d\sigma^-}=-{N\over 48}\bigl(
1-(1+P_\infty e^{\sigma^-})^{-2}\bigr) \,,
\ee
which is indeed proportional to the number of matter fields.
We have successfully removed the non-propagating modes
from the Hawking radiation at the semiclassical level.  There is,
however, a potential conflict here with the no-ghost
theorem,\footnote{See \cite{verl} for a discussion of the
no-ghost theorem in two-dimensional dilaton gravity.}
which has not been addressed.

\subsection{Conformally Invariant Boundary Conditions}
\label{fourfour}
A closer examination of the expression (\ref{bondi2}) reveals
a disastrous instability.  If we evaluate it at
late retarded times, $\sigma^-\rightarrow +\infty$, we find
that $m(\sigma^-)$ goes to minus infinity.  The vacuum is
unstable under arbitrarily small perturbations.  This is of
course the same disaster we encountered previously in the
RST model.  In that case it was avoided by restricting the
range of the $\Omega$ field variable and imposing the RST
boundary conditions at $\Omega=\omcr$.

We could adopt the same strategy here but there is a price to
pay.  If the range of $\Omega$ is restricted the functional
integral does not define a conventional
quantum field theory and it is unknown how to carry it out in
a manner consistent with general covariance.  We did not
worry about this problem in Section~\ref{seciii} since we were
only trying to solve semiclassical equations but it would have
to be faced if we wanted to carry out a quantization of the
RST model.  Furthermore, as we noted before, the boundary
conditions (\ref{rstbc}) on $\Omega$ appear to be incompatible
with simple quantum mechanical boundary conditions on the
matter fields.

An alternate procedure \cite{chuver,dasmuk,aslt},
is to place no restriction on the
field values of $Y$ but impose boundary conditions on all
fields in the theory along a timelike curve $x^+x^-=constant$.
We can interpret this curve as the origin of radial coordinates
$r=0$ from the point of view of a higher-dimensional theory.
If we define $s^\pm=\pm \log{x^\pm}$ then the boundary curve
$x^+x^-=constant$ is at a constant value of the spatial
coordinate $s^1=s^1_0$.

Our boundary conditions need to satisfy a number of
physical requirements.  First of all, they should be
consistent with the conformal symmetry of the bulk theory in
order to ensure general covariance and energy conservation.
Our second requirement is that the vacuum configuration of
the bulk theory, or a close cousin, must be compatible with
the boundary conditions.  If the nature of the vacuum is
greatly altered it can be difficult to interpret the physics
of the model in terms of black holes.  Finally, the boundary
conditions should effect a cure of the negative energy
instability of the bulk theory.  We will be able to achieve
stability under small perturbations,
which is enough to allow the perturbative construction of a
low-energy $S$-matrix for asymptotic observers.
Large incoming pulses will, however, still produce low values
of $Y$ and destabilize the system.

Our analysis will be semiclassical in that we treat $X$ and
$Y$ as c-number fields when we impose the boundary conditions.
It is presumably possible to modify the boundary conditions
order by order in the loop expansion to maintain conformal
invariance, or even define an exact boundary conformal field
theory, but this is beyond the scope of our discussion.

The semiclassical approximation is not always reliable, but
corrections to it can be systematically suppressed by taking
the large $\gamma$, \ie\ large $N$, limit.
In taking this limit $2\tilde X=2X-\log{\gamma}$ and
$\tilde T^f_{\pm\pm}= {1\over \gamma}T^f_{\pm\pm}$ are held
fixed.  The gravitational part of the semiclassical action
(\ref{xyact}) becomes
\beq
\label{gpart}
\lefteqn{
S_g={\gamma\over \pi}\int d^2y \bigl[
-\partial_+\tilde X\partial_-\tilde X
+\partial_+Y\partial_-Y
}\nonumber \\
& & \>
+e^{2(\tilde X-Y)} \bigr] \,,
\eeq
and the constraints can be written
\be
\partial_\pm Y \partial_\pm Y
-\partial_\pm \tilde X \partial_\pm \tilde X
+\partial_\pm\partial_\pm \tilde X
+\tilde T^f_{\pm\pm} =0 \,.
\label{Nconstr}
\ee
In $s$-coordinates the vacuum solution of the bulk theory
takes the simple form
\be
\!\!\begin{array}{l}
Y = \exp{(2s^1)} - {1\over 2}\,s^1 \,,
\\ { } \\
\tilde X = \exp{(2s^1)} + {1\over 2}\,s^1 \,.
\end{array}
\label{bulkvac}
\ee

Since the boundary is a straight
line in this coordinate system, boundary conformal invariance
is equivalent to \cite{cardy}
\be
\tilde T_{++}(s^0,s^1_0)=\tilde T_{--}(s^0,s^1_0) \,,
\label{ttpm}
\ee
where $\tilde T$ is the total energy-momentum tensor for all
fields. If either Dirichlet or Neumann boundary conditions are
imposed on the matter fields,
\be
\partial_+f_i(s^0,s^1_0) \pm \partial_-f_i(s^0,s^1_0)=0 \,,
\label{dnff}
\ee
then the boundary condition (\ref{ttpm}) implies
\be
\!\!\begin{array}{l}
\partial^2_+\tilde X-\partial_+\tilde X\partial_+\tilde X
+\partial_+Y\partial_+Y \\ { } \\
= \partial_-^2\tilde X-\partial_-\tilde X\partial_-\tilde X
+\partial_-Y\partial_-Y \,.
\end{array}
\label{xxyy}
\ee
A simple semiclassical solution of this equation is given by
\be
\!\!\begin{array}{l}
Y=Y_0 \,, \\ { } \\
\partial_+\tilde X - \partial_-\tilde X= Ae^{\tilde X-Y_0} \,,
\end{array}
\label{bctw}
\ee
where $A$ and $Y_0$ are constants.  More general solutions can
be found but we will only consider this example here.

Insisting that the vacuum solution (\ref{bulkvac})
be compatible with the boundary conditions
determines the parameters $A$ and $Y_0$ in terms of
$s^1_0$, the location of the boundary.  The constraint on $Y_0$
follows directly from the $Y$ boundary condition in (\ref{bctw}),
\be
Y_0 = \exp{(s^1_0)} - {1\over 2}s^1_0 \,,
\label{tmsn}
\ee
and the $\tilde X$ boundary condition is satisfied by the vacuum
solution only if
\be
A  =
2 \cosh{(s^1_0 + \log{2})} \,.
\label{tmln}
\ee
Note that $A\geq 2$ for all values of $s^1_0$, which means that
vacuum compatibility does not allow the theory to be analyzed
perturbatively in the strength of the boundary
interaction.\footnote{A=0 is an allowed value in the special
case of $\gamma=0$ \cite{dasmuk,verl}.}
For a given allowed value of $A$ there are two values of $s^1_0$
consistent with (\ref{tmln}), but it turns out that
only one of them makes the vacuum stable under small incoming
energy perturbations.

\subsection{A Dynamical Boundary Curve}
\label{fourfive}
A boundary condition imposed at fixed $s^1$ restricts the left and
right conformal invariance to a diagonal subgroup. Separate left and
right invariance can be regained, however, at the price of allowing
the boundary to follow a general trajectory, described by an equation
$y_b^-(y^+)=y^-$.  Holding $Y$ fixed along the boundary implies
\be
u^{-1}\partial_+Y+u\, \partial_-Y=0 \,,
\label{ybcg}
\ee
where
\be
u(y^+)\equiv
\bigl(\frac{\partial y_b^-}{\partial y^+}
\bigr)^{1/2} \,,
 \label{udfn}
\ee
and $(u^{-1},u)$ is a tangent vector to the boundary
curve in the $y$-coordinate system.  Neumann or Dirichlet
conditions on the matter fields become
\be
u^{-1}\partial_+f_i\pm u\,\partial_-f_i=0 \,,
 \label{fbcg}
\ee
for a general boundary curve and the boundary condition on $X$
in (\ref{bctw}) becomes
\be
u^{-1}\partial_+X- u\,\partial_-X
= Ae^{X-Y_0}+u^{-2}\partial_+u \,.
\label{xbcg}
\ee
The last term arises because $X$ contains the conformal factor
and does not transform as a scalar under a change of coordinates.

By acting on both sides of (\ref{xbcg}) with the operator
$u^{-1}\partial_++u\partial_-$, which generates translations
along the boundary, we obtain
\be
\!\!\begin{array}{l}
u^{-2}(\partial_+^2X-\partial_+X\partial_+X)
\\ { } \\
= u^2 (\partial^2_-X - \partial_-X\partial_-X)
- u^{-1}\partial_+^2(u^{-1})\,.
\end{array}
 \label{xnml}
\ee
This identity, together with the $Y$ boundary condition (\ref{ybcg}),
can be used to relate the components of the gravitational stress
tensor at a general boundary curve,
that
\be
u^{-2}\tilde T^g_{++}=u^2\tilde T^g_{--}-
u^{-1}\partial_+^2(u^{-1}) \,.
 \label{tgpm}
\ee
The last term is familiar from the study of moving mirrors
\cite{mirrors}. It vanishes in ``straight line" gauges for which $u$
is constant.

The dynamics of the boundary is governed by an ordinary differential
equation.  We find it convenient to derive it in Kruskal gauge,
where $X=Y$ and the boundary conditions (\ref{ybcg}) and (\ref{xbcg})
can be combined into
\be
2u^{-1} \partial_+ Y
=  A+u^{-2}\partial_+ u \,.
\label{wtfd2}
\ee
Multiplying by $u$ and then differentiating along the boundary, one
obtains
\be
2\partial_+^2 Y+2u^2\partial_-\partial_+ Y =  A \partial_+u
+\partial_+^2 \log{u} \,.
 \label{nggw2}
\ee
The following relations hold for a general solution of the
equations of motion in Kruskal gauge,
\be
\!\!\begin{array}{l}
\partial_+Y = -
\bigl(x^-+\tilde P_++\frac{1}{4x^+}\bigr) \,,
\\ { } \\
\partial^2_+ Y = - \tilde T^f_{++}+\frac{1}{4x^{+2}} \,,
\\ { } \\
\partial_-\partial_+ Y = -1 \,.
\end{array}
\label{kgxq2}
\ee
Substituting into (\ref{nggw2}) leads to
\be
\partial_+^2 \log{u} + A {\partial_+u}
+2u^2 -\frac{1}{2 {x^+}^2}
=-2 \tilde T^f_{++}.
\label{fqnw}
\ee
An alternate form of this boundary differential equation is
obtained by transforming to the $s$-coordinate system and
defining $\omega(s^+)=s^+ + \log{u}$.  This change of variables
is useful because $\omega$, unlike $u$, is a constant in the
vacuum.  The resulting equation
can be given an interpretation in terms of a particle moving
in a potential subject to a driving force and a non-linear
damping force,
\be
\omega'' + k(\omega)\omega'
+\frac{\partial V(\omega)}{\partial \omega}
=-2\tilde T^f_{++}\,.
\label{wwqn}
\ee
The primes denote differentiation with respect to $s^+$
while $k(\omega)=A\exp{(\omega)} -1$ and
$\frac{\partial V(\omega)}{\partial \omega}
=2\exp{(2\omega)}-A\exp{(\omega)}
+\frac{1}{2}$.
The damping arises because boundary energy can be
dissipated into (or absorbed from) the rest of the spacetime,
and it becomes negative for sufficiently negative $\omega$.

\subsection{Vacuum stability}
\label{foursix}
Given that $\tilde T^f_{++}=0$ and $\omega$ is a constant in the
vacuum the boundary equation (\ref{wwqn}) implies that the vacuum
corresponds to the particle sitting at an extremum of its
potential.
There is a local minimum at
$\omega=\omega_0$, where
\be
\exp{(\omega_0)} = {1\over 4}
(A+\sqrt{A^2-4}) \,,
\label{locmin}
\ee
Since the damping in
the boundary equation can be negative it is not sufficient to
identify a local minimum of the potential to ensure stability
under small perturbations, but it is straightforward to
linearize (\ref{wwqn}) around the vacuum solution and carry
out a stability analysis \cite{aslt}.  One finds that the
system will return to the vacuum after being excited by a
weak pulse of incoming energy for all $A>2$.

The stability conditions can also be expressed in terms of
the parameter $Y_0$.  They reduce to $Y_0>Y_{min}$ where
$Y_{min}$ is the value of $Y$ where the field redefinition
(\ref{xydef}) is degenerate.  The boundary conditions can only
stabilize the system under small perturbations when the boundary
is placed on the physical side of $Y_{min}$.

In summary, there exist conformally invariant boundary
conditions, imposed at a timelike boundary placed on the
weak coupling side of $Y=Y_{min}$, which ensure that weak
pulses incoming from ${\cal I}^-$ are reflected to ${\cal I}^+$,
although in a distorted form.  This allows in principle the
construction of an $S$-matrix for low-energy asymptotic
observers in a perturbative expansion in the strength of the
incoming pulses.

The behavior for large pulses is quite different and this has
important implications for the utility of models of this type
for black hole physics.  Consider a pulse which begins at
$x^+=x^+_i$ in Kruskal coordinates and carries total Kruskal
momentum $\tilde P_\infty$.  We saw previously
that in the absence of a boundary the Bondi mass goes to minus
infinity at a point on ${\cal I}^+$, which is located at
$x^-=-\tilde P_\infty$.  One might expect this behavior to change
in the presence of a boundary and we have seen that this is indeed
the case for weak incoming pulses.  There is, however, a general
argument stating that boundary conditions cannot stabilize the
evolution when a large pulse is incident on the vacuum.
The pulse first reaches the boundary at
\be
(x^+,x^-)=(x^+_i,-\exp{(2s^1_0)}/ x^+_i) \,.
\label{firsthit}
\ee
By causality the behavior on ${\cal I}^+$ cannot be influenced
by boundary reflection prior to $x^-=-\exp{(2s^1_0)}/x^+_i$,
and therefore the Bondi mass will still plunge to minus infinity
at $x^-=-\tilde P_\infty$ if
\be
\tilde P_\infty > \exp{(2s^1_0)}/ x^+_i \,.
\label{largep}
\ee
Given any local boundary conditions there is always a sufficiently
large incoming momentum for which the disaster occurs.

Since the
causal past of ${\cal I}^+$ may include only regions of weakly
coupled dynamics this instability cannot in general be averted by
just changing the strongly coupled dynamics of the model.
It appears to require fundamentally new input, such as the
endpoint prescription in the RST model which in effect
involves topology change but it remains an unsolved problem to
incorporate such processes consistently into the quantum
theory without encountering the negative energy instability
\cite{strominger,jpas,lpst}.

It should be noted that the above disaster does not imply
a sickness in the $X,Y$ conformal field theory itself.  It arises
in the transcription from the $X,Y$ conformal field theory to
a $\rho,\phi$ theory of dilaton gravity.  The point
$x^-=-P_\infty$ is at a finite distance in the fiducial metric used to
regulate the $X,Y$ conformal field theory, and the $X$ and $Y$
fields can be continued past this point. The reflected pulse, and the
information it carries, eventually comes back out.  This, however,
occurs ``after the end of time" as measured by the physical metric,
which has conformal factor $\rho$.

The negative energy instability is linked to the infinite specific
heat of black holes in these models in that the rate of Hawking
evaporation does not change as the energy of the black hole is
depleted.  A possible way around this problem is to include
gauge fields in the theory so that it has two-dimensional
analogs of charged black holes \cite{olcnvf}, or consider
Reissner-Nordstrom black holes in the spherically symmetric
approximation.
In the extremal limit $M=Q$ the black hole temperature goes to zero
and it has been argued that an extremal black hole is a stable
endpoint of Hawking evaporation in such models \cite{charged}.
Unfortunately the known toy models with
charged black holes are much less amenable to analytical study
than the CGHS model and its cousins, but perhaps simpler models
can be developed.

\section{Black Hole Complementarity}
\label{secv}
In the preceding sections we took a semiclassical approach to
black hole physics and the information problem.  The starting
point was a classical theory of gravity coupled to matter fields
and then we included quantum effects and their back-reaction
on the geometry in a series of steps designed to successively
capture more features of the full quantum theory.
Our discussion was in the context of two-dimensional
toy models but the general philosophy of the semiclassical
approach is in essence the same there as in the more challenging
four-dimensional theory.

\subsection{Information Lost}
\label{fiveone}
The fate of quantum information is
decided at the event horizon, which, for a large mass black hole,
is in a region where the spacetime curvature and other local
coordinate invariant features of the geometry are weak.
The semiclassical approach assumes that it follows from the
above that only low-energy effects are involved in deciding
the issue, and we are justified in using a local effective field
theory to describe the physics.

It should be noted that a fully consistent effective field theory
of black hole evolution has yet to be
constructed.  The two-dimensional models we have considered here
are afflicted with instabilities at the quantum level and such
technical issues are even less under control in higher-dimensional
theories.  Further work may, however, lead to more successful models and
let us assume, for the moment, that a local effective description
of gravity at low energies can be found.

Imagine a team of very patient, technologically advanced observers
studying the formation of a black hole and
its subsequent evaporation from a safe distance.  The observers
initially prepare a pure quantum state of infalling matter and then
make careful measurements on the Hawking radiation emitted by the
black hole over its entire lifetime.  To determine whether the
final state is mixed or pure, our observers will have to perform
an enormous number of such experiments, using
identically prepared initial states, because only mutually commuting
observables can be measured in any single run.  They also have to be
able to make sophisticated observations of correlations between quanta
emitted at different times in the life of the black hole, for
even if the formation and evaporation process as a whole is
governed by a unitary $S$-matrix the radiation emitted at any given
moment will appear thermal.

The question is whether the infalling matter will give up all
information about its quantum state to the outgoing Hawking
radiation or whether the information gets carried into the black
hole.  If the information is imprinted on the Hawking radiation
then it must also be removed from the infalling matter as it
approaches the event horizon,\footnote{By Hawking radiation we only
mean the radiation that carries off the energy of the black hole
during its evolution and not any slow radiation that might emanate
from a Planck scale remnant.} for otherwise we would have a duplication
of information in the quantum state in violation of linear quantum
mechanics \cite{stu}.

A book set on fire is a useful analogy.  All the information initially
contained on the pages can in principle be gleaned from measurements on
the outgoing smoke and radiation, but at the end of the day this
information is no longer available in book form.

There is a crucial difference, however, between a burning book and
matter falling into a black hole.  In the former case it is a well
understood microphysical process which transfers the information from
book to radiation, whereas matter in free fall entering a black hole
encounters nothing out of the ordinary upon crossing the event horizon.
If both the infalling matter and the geometry well away from the black
hole singularity are described by a local, weakly coupled, low-energy
effective field
theory, then the question of whether an observer passes unharmed
through the horizon or is disrupted before entering the black hole
appears to have a coordinate invariant answer.
Clearly, no disaster happens at the horizon
in the local free fall frame and one concludes that
information is not carried out in the Hawking radiation.

\subsection{Information Regained}
\label{fivetwo}
At this point there are different roads to go by.  One of them
is to accept the premise of the above argument and resign oneself to
the information going into the black hole.  The task at hand is
then either to come up with a consistent theory where quantum
information is lost or a theory where unitarity is maintained by
having long-lived black hole remnants.

We will not follow this road here, but instead
{\it postulate\/} that all information about the initial quantum
state of infalling matter forming a black hole is returned to
outside observers and is encoded in the outgoing Hawking radiation
as the black hole radiates.  In this view there is no fundamental
information loss and any stable or long-lived black hole remnants
are finitely degenerate at the Planck scale.  This is a conservative
viewpoint in that it assumes unitarity in all quantum
processes, even when gravitational effects are taken into account,
but, as we shall see, it presents a novel view of spacetime physics
near an event horizon.  This approach was pioneered by Page
\cite{page1} and `t~Hooft \cite{thooft1} and has been advocated by a
growing number of authors in recent years, including
\cite{stu,verl,maggiore,mathur}.

The remainder of these lecture notes will be concerned with
exploring general consequences of the above postulate.
We are first of all led to conclude that the physical description
of matter approaching the event horizon differs between the
asymptotic and free fall reference frames by something more than is
warranted by the usual behavior of local fields under coordinate
transformations.  How serious is this apparent contradiction?
The gravitational redshift between the two frames is enormous; the
relative boost factor grows exponentially with the time measured
in the asymptotic frame and, as 't~Hooft has emphasized
\cite{thooft1}, it becomes much larger than anything that has been
achieved in experiments.  It is therefore legitimate to question
whether the usual Lorentz transformation properties of localized
objects correctly relate observations made in the two frames
\cite{susskind}.

The principle of black hole complementarity \cite{stu} states that
there is no contradiction between outside observers finding
information encoded in Hawking radiation, and having observers
in free fall pass unharmed into a black hole.
The validity of this principle rests on matter having unusual
kinematic properties at high energy but we will argue that it does
not conflict with known low-energy physics.  The basic point is that
the apparent contradiction only comes about when we attempt to
compare the physical description in different reference frames.
The laws of nature are the same in each frame and low-energy
observers in any single frame cannot establish duplication of
information.  This will be illustrated below with the help of some
gedanken experiments involving black holes but let us first consider
the physical picture presented in the outside frame in a little more
detail.

\subsection{The Stretched Horizon}
\label{fivethree}
It is well established that, from the point of view of outside
observers, the classical physics of a quasistationary black hole
can be described in terms of a `stretched horizon' which is a
membrane placed near the event horizon and endowed with
certain mechanical, electrical and thermal properties
\cite{mempar}.
The nature of this description is coarse grained in that it is
dissipative and irreversible in time.  One doesn't have to be very
specific about how near the event horizon the stretched horizon is
placed as long as it is close compared to the typical length scale
of the classical problem, which could for example be to describe a
black hole interacting with a companion in a binary.

Susskind, Thorlacius, and Uglum \cite{stu} proposed to
go beyond this classical picture by postulating that the coarse
grained thermodynamic description of the classical theory has an
underlying microphysical basis.  These authors did not specify at
the time what the nature of this microphysics is, but Susskind
\cite{susskind} subsequently pointed out that relativistic strings
exhibit precisely the kinematic behavior that is required for black
hole complementarity to hold and argued that the microphysics of
the stretched horizon should be understood in the context of string
theory.  We will discuss these arguments in Section~\ref{secvii}
but for now we will proceed with a phenomenological description of
black holes in terms of a quantum mechanical stretched horizon,
which is a membrane, carrying microphysical degrees of freedom,
with an area larger than that of the event horizon by one Planck unit.
The term Planck unit is being used in a loose sense and
simply refers to the high-energy scale at which the radical kinematic
behavior, required for returning the information, enters.  In string
theory this would be the fundamental string scale
which can be considerably lower than the usual Planck energy.
In addition, to implement black hole complementarity we have to
stipulate that the membrane has no substance in the
frame of an observer entering the black hole in free fall.

The evaporation of a large black hole is a slow process and for
many purposes the evolving geometry is well approximated by a
static classical solution.  We will only be discussing non-rotating,
electromagnetically neutral black holes, for which the
Schwarzschild line element is
\be
ds^2=-(1{-}{2M\over r})dt^2+(1{-}{2M\over r})^{-1}dr^2
+r^2d\Omega^2 .
\label{schwarzschild}
\ee
An outside observer who is at rest with respect to the
Schwarzschild coordinate system sees thermal radiation at a
temperature which depends on the spatial position,
\be
T(r)={1\over 8\pi M}\bigl(1-{2M\over r}\bigr)^{-1/2} \,.
\label{stattemp}
\ee
Near the black hole this temperature goes like
$T\approx (2\pi\delta)^{-1}$, where $\delta$ is the proper
distance between the observer and the event horizon.  The high
temperature radiation can be attributed to the acceleration required
to prevent the observer from falling into the black hole, which
diverges in the $\delta\rightarrow 0$ limit.

In our phenomenological approach the region nearest the event
horizon, where the temperature (\ref{stattemp}) is diverging,
will be replaced by a hot membrane placed at a proper distance
of one Planck unit outside the event horizon, which corresponds
to a stretched horizon area of order one larger than the area of the
event horizon.

As far as outside observers are concerned the
stretched horizon can be viewed as the source of Hawking radiation.
As a surface at Planckian temperature it emits particles copiously
but most of these do not have sufficient energy to escape the
gravitational pull of the black hole.  Those who do are predominantly
in a low angular momentum channel and are redshifted to energies of
order the Hawking temperature when they reach the asymptotic
region \cite{stu,mempar}.

\includefigures\begin{figure}[bt]
\epsfxsize=6.cm \epsfysize=6cm
\epsfbox{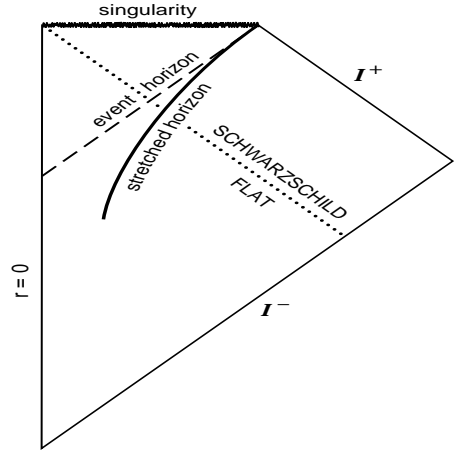}
\caption{Penrose diagram for a black hole formed by an infalling
shell of lightlike matter.}
\label{match}
\end{figure}\fi

Now consider a nonstatic geometry corresponding to the formation
of a large black hole by infalling matter.  The simplest case to
study involves a thin spherical shell of massless matter
\cite{shell}.  The geometry is constructed by matching flat spacetime
and a massive Schwarzschild solution across a radial null surface
as shown in Figure~\ref{match}.  The solution for a more general
distribution of incoming null matter can approximated by a sequence
of shells with the appropriate matching conditions satisfied at each
shell.

A stretched horizon can be defined as a timelike surface just
outside the event horizon.  There are a number of ways to achieve
this, but a simple method is as follows \cite{stu}.  At a point
on the global event horizon construct the past directed radial null
ray which does not lie in the horizon itself.  The stretched horizon
is defined to intersect this ray at a point where the area of
the transverse two-sphere has increased by an amount of order a
Planck unit relative to the area where the ray intersects the event
horizon.  A feature of the collapse geometry as compared
with a static black hole is that both the event horizon and the stretched
horizon extend into the flat spacetime region inside the infalling matter
shell.  With our definition the stretched horizon begins at the same
finite value of advanced time as the event horizon itself
(see Figure~\ref{match}).  At this point the area of the stretched
horizon is only one Planck unit and it is not very meaningful to
extend its definition to earlier times.

The stretched horizon defined in this way has the important property
that the proper acceleration is of order one in Planck units everywhere
on it, independent of time, even including the point where it crosses
the incoming matter shell.  As a result the stretched horizon can again
be viewed as a hot boundary with a Planckian temperature.

\section{Gedanken Experiments Involving Black Holes}
\label{secvi}
In this section we will illustrate the concept of black hole
complementarity by considering some gedanken experiments where
one might expect contradictions to arise.  The main conclusion
will be that apparent contradictions can always be traced to
unsubstantiated assumptions about physics at or above the
Planck scale.\footnote{The relevance of short distance physics
to the information puzzle was also emphasized in the work of
't~Hooft \cite{thooft1} and of Schoutens, Verlinde, and Verlinde
\cite{verl}.}
This observation does not resolve the information
problem but it challenges the commonly held view that the
paradox can be posed without any reference to the
underlying short-distance physics.

\subsection{A Test of Information Duplication}
\label{sixone}
It is important to determine whether black hole complementarity
leads to observable duplication of information.
The results of measurements performed inside a black hole are not
available to outside observers.  Consider therefore a gedanken
experiment \cite{lslt2} where an
observer first learns the result of a measurement made on the
outgoing Hawking radiation and then enters the black hole in
order to receive a signal from some system that previously fell
through the event horizon.  The system could for example be a
measuring apparatus $A$ which carries one member of a pair of
`spins' $a$ and $b$, that have been prepared in a singlet state
outside the black hole.
By `spin' we mean an internal label which is not coupled to a
long-range gauge or gravitational field so that it does not
lead to classical black hole hair.
The apparatus is programmed to measure the spin $a$ once it is
inside the event horizon and transmit the result.
The other member of the spin pair $b$ remains outside and
a team of observers, hovering at a safe distance
outside, makes measurements on the Hawking radiation.  Since all
infalling information is assumed to be radiated out the observers
will eventually be able to determine the state of the
spin that was carried inside.\footnote{In order to measure
statistical quantum correlations the experimenters would work with
an ensemble of identically prepared spin pairs.}
Armed with this information one of the observers $O$ now enters
the black hole in order to receive the signal from the apparatus.

\includefigures\begin{figure}[bt]
\epsfxsize=7.5cm \epsfysize=7cm
\epsfbox{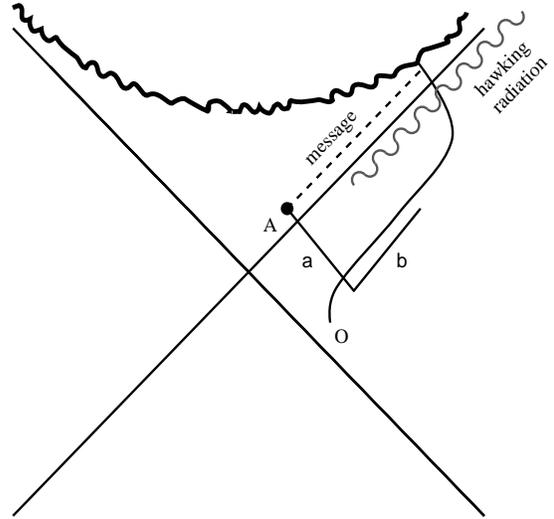}
\caption{A gedanken experiment designed to test duplication of
quantum information.}
\label{gedank1}
\end{figure}\fi

It would appear that our observer has managed to learn the results
of two separate measurements of the same spin.  This violates
the principles of quantum mechanics as can be seen as follows.
The spin $a$ that is carried into the black hole is anti-aligned
with $b$, the one that remains outside.
{}From the point of view of an external observer a
measurement of the `spin in the Hawking radiation' $h$ amounts to
a measurement of the original spin $a$, and therefore $h$ must
also be anti-aligned with $b$.  The observer $O$ would learn that
measurements of $a$ and $h$ along any axis would always give the
same result, while for any quantum state of two spins such
measurements can at most agree along a single axis.

It turns out to be impossible to carry out this experiment
employing only low energy physics \cite{jphv}.
The outside observers have to carry out correlation measurements
on the outgoing Hawking radiation for a very long time before they
can hope to recover the information about the spin that is sent
into the black hole.
This time was estimated in \cite{lslt2} to be a finite fraction
of the black hole lifetime, \ie\ of order $M^3$.
The estimate is based on some general results found by Page
\cite{page2} on the entanglement entropy of subsystems in
statistical mechanics.  Based on this, let us take as given that
the distant observers have to conduct experiments for a period of
order $M^3$ before any information can be recovered.
During this time the mass of an evaporating black hole
significantly decreases but we shall ignore this and make use
of a static Schwarzschild geometry.  It will become apparent that
the effect of evaporation is to strengthen the conclusion from
the static case.

Since we want to discuss measurements made inside a black hole it
is convenient to use Kruskal coordinates $U$ and $V$ which extend
past the event horizon and are defined in terms of Schwarzschild
coordinates through
\be
\!\!\begin{array}{l}
-UV=16M^2({r\over 2M}-1)\exp{({r\over 2M}-1)} \,,
\\ { } \\
-U/V=\exp{({t\over 2M})} \,.
\end{array}
\label{kruskals}
\ee
The Schwarzschild line element (\ref{schwarzschild}) becomes
\be
ds^2 = -{2M\over r}\exp{(-{r\over 2M}{+}1)} dUdV
+r^2d\Omega^2.
\label{klineel}
\ee
The future event horizon is at $U=0$ and the singularity is at
$UV=1$.  The geometry is shown in Figure~\ref{gedank1}.
If we choose the origin of our time coordinate so that the
apparatus passes through the event horizon at $V=1$ then the
observer $O$, who waits outside for a period of order $M^3$ as
measured by Schwarzschild time, will enter at
$V\sim\exp{(M^2)}$.  If the message is to be received before
$O$ hits the singularity it must be sent before the apparatus
reaches $U\sim\exp{(-M^2)}$.   Near $V=1$ this corresponds to
an extremely short proper time $\tau\sim M^2\exp{(-M^2)}$, and
the uncertainty principle then dictates that the message must
be encoded into radiation with super-Planckian frequency
$\omega\sim M^{-2}\exp{(M^2)}$.  It is clear that the
apparatus cannot communicate the result of its measurement to
the observer $O$ employing only low-energy physics.  The
evaporation of the black hole only makes the time available
for $O$ to receive the message shorter.

It is a generic feature of gedanken experiments of this type that
short distance physics enters into their analysis in an essential
way and any apparent contradiction with black hole
complementarity can be traced to unwarranted assumptions about
physics beyond the Planck scale.

\includefigures\begin{figure}[bt]
\epsfxsize=7.5cm \epsfysize=7cm
\epsfbox{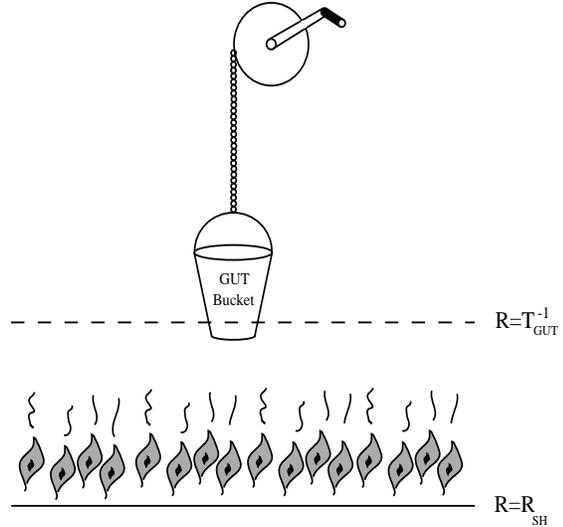}
\caption{A gedanken experiment to test baryon number non-conservation
near the stretched horizon.  $R$ denotes the proper distance from
the event horizon.}
\label{gedank2}
\end{figure}\fi

\subsection{Thermal Effects Near the Stretched Horizon}
\label{sixtwo}
Another class of experiments involves attempts by external
observers to detect whether quantum information is stored at the
stretched horizon.  Their analysis also requires short-distance
physics even for a large black hole \cite{lslt2}.  We will not
discuss those
experiments here but a closely related one instead, which illustrates
how knowledge of high-energy effects can help resolve an apparent
contradiction between observations in different reference frames.
The effect in question is the violation of baryon number in the
process of black hole formation and evaporation.  The high-energy
physics involved is at the GUT scale rather than the Planck scale.

\def\mgut{M_{\rm GUT}}
One of the thermal effects we expect to take place in the reference
frame of asymptotic observers is baryon number violation.
This effect will become noticeable when matter passes within a proper
distance of order $\mgut^{-1}$ of the event horizon and will not be
hidden inside the black hole.  In order to verify this, a
technologically advanced observer
could prepare a sealed `bucket' with walls that
transmit heat but not baryon number.  The bucket is then slowly
lowered towards the black hole and retrieved after its bottom edge
comes within a proper distance of $\mgut^{-1}$ from the event
horizon, as indicated in Figure~\ref{gedank2}.
The baryon number in the bucket is measured before and
after lowering and according to the asymptotic observer thermal
effects induce a change.
The effect can be attributed to the intense acceleration,
which the bucket undergoes to avoid falling into the black hole.

Now imagine that the bucket is dropped into the black hole.
The asymptotic observer would again say that the bucket encounters
a thermal bath but an inertial observer traveling with the bucket
in free fall would disagree.  This does not lead to any contradiction
unless the infalling observer manages to count the baryons in the
bucket and transmit the result to the outside observer before the
bucket passes through the event horizon.  The bucket will, however,
only spend a time of order $\mgut^{-1}$ in the region of interest,
according to the clock in the free fall frame, and the infalling
observer must complete the baryon count in that time.  This is
precisely the timescale of short lived processes which continuously
violate baryon number in the bucket rest frame.  Such rapid counting
of baryons will not yield the same answer as a leisurely measurement
at the beginning of the experiment and the contradiction is avoided.

\section{The Stretched Horizon in String Theory}
\label{secvii}
If the ideas presented above are correct then it is essential to
gain understanding of physics at very short distances in order to
fully resolve the issue of information loss.  String theory is
widely believed to provide a consistent short-distance description
of matter and gravity and Susskind \cite{susskind}
has argued that the kinematic behavior of fundamental strings
is entirely consistent with the requirements of black hole
complementarity.  The basis for this claim is
that zero-point fluctuations of string modes make the
size of a string depend on the time resolution employed \cite{kks}.
The shorter the time over which the oscillations of a string are
averaged the larger is its spatial extent will appear.

\subsection{Infalling String Near The Horizon}
\label{sevenone}
Consider a string configuration in free fall approaching a
black hole event horizon.  An observer at rest far away from
the black hole measures asymptotic time, but because of the
increasing redshift, a unit of asymptotic time corresponds to
an ever shorter time interval in the free-fall frame.  The
distant observer is therefore using a shorter and
shorter resolution time to describe the string configuration
and, once it passes within a proper distance of order the
string scale from the event horizon, the string begins to
spread both in the longitudinal and transverse directions.
The longitudinal spread is sufficiently rapid to
cancel out the longitudinal Lorentz contraction caused by the
black hole geometry.
Meanwhile, the spread in the transverse directions causes the
configuration to cover the entire horizon area in a time
which is short compared to the black hole lifetime.

The stretched horizon is thus {\it made out of\/} the strings
in the infalling matter which forms the black hole.
On the other hand, this spreading effect is not present
in the free-fall frame, where there is no redshift
to enhance the time resolution, and from the point of view
of an infalling observer there is no stretched horizon, in line
with the principle of black hole complementarity.

The string spreading effects in which we are interested take
place on a short timescale compared to the black hole lifetime
\cite{susskind} and we therefore consider a static classical
geometry.  Furthermore, the spreading takes place in a thin
layer, whose proper thickness is of order the string scale
$1/gM_{Pl}$, outside the event horizon and for a
macroscopic black hole this region is well approximated
by Rindler space.  This is apparent if we consider the
the Schwarzschild line element, expressed in Kruskal coordinates,
as $r\rightarrow 2M$.   In this limit the event horizon looks
locally like a planar surface and (\ref{klineel}) reduces to
\be
ds^2 = - dU\,dV + d{X^\perp}^2 \,.
\label{rlineel}
\ee
The null coordinates $U,V$ extend into the black hole interior
and are appropriate for an observer passing through the horizon
in free fall.  They are related to the null coordinates $u,v$ of
an asymptotic observer through,
\begin{equation}
{U\over 4M}=-e^{-u/4M}\,,\qquad
{V\over 4M}=e^{v/4M} \,.
\end{equation}
The usual description of Rindler space is obtained by rescaling
all the coordinates to absorb the factors of $4M$, but here we will
keep these factors explicit.

Rindler space is isomorphic to a slice of flat Minkowski space,
where we can discuss free string propagation in light-cone
gauge.  Consider an infalling string described by transverse
coordinates $X^i(\sigma,\tau)$, with $i=1,2$, and some
internal degrees of
freedom depending on the string model in question.
In the free-fall frame the light-cone gauge condition is
$\tau=U/4M$.  The internal degrees of freedom decouple from the
transverse coordinates which satisfy a free wave equation,
\begin{equation}
\label{waveeq}
\bigl[ \partial^2_\tau -\partial^2_\sigma \bigr]
X^i (\sigma,\tau) = 0 \,.
\end{equation}
The solution can be expressed by the usual sum over modes of
oscillation, but, unless the infinite sum is cut off in some way,
this leads to ill-defined expressions for quantities such as the
average transverse area occupied by the string in the quantum
theory \cite{kks}.
Introducing a cutoff on the mode expansion corresponds
physically to employing finite time resolution so that mode
oscillations above a given frequency average out.
The string wave-function extends over a transverse area which
grows logarithmically with better resolution time while the
average length of string projected onto the transverse
directions grows linearly.  This means that the string
density will increase at the center of the distribution, and
eventually we can no longer neglect the effect of string
interactions there.

Now we want to consider the string evolution in the reference
frame of a distant fiducial observer, whose retarded time,
$t=u$, is related to the worldsheet parameter time through
\begin{equation}\label{timerel}
{\tau} = - e^{-{t / 4M}} \>.
\end{equation}
As asymptotic time goes on, a fixed resolution in $t$ thus
corresponds to an exponentially improving resolution in $\tau$,
so that in the reference frame of the distant observer, the string
wave-function will spread with time to occupy a transverse
area proportional to $t$.
By using the constraint equations of light-cone string theory
one can show that the string also spreads in the longitudinal
direction and that the longitudinal spreading is rapid enough
to balance the Lorentz contraction due to the black hole
metric near the event horizon \cite{susskind}.

\subsection{Branching Diffusion of String Bits}
\label{seventwo}
Since the stringy stretched horizon is formed from the infalling
matter itself, it efficiently absorbs the quantum information
contained in that matter.  The string spreading process also
thermalizes the stretched horizon \cite{mpt}.  This is seen as
follows.
In asymptotic time $t$ the equation of motion (\ref{waveeq}) of
the field $X^i\bigl(\sigma,\tau(t)\bigr)$ becomes
\begin{equation}
\label{tdeforx}
\bigl[\partial_ t^2+
\frac{1}{4 M}\partial_t -
\bigl(\frac{e^{-t/4M}}{4M} \bigr)^2
\partial_\sigma^2 \bigr]
X^i (\sigma,t) = 0 \,.
\end{equation}
Now split both the field $X^i$ and its conjugate momentum
$\partial_tX^i$ into a slowly varying, classical part and
fast varying, quantum part,
\be
\!\!\begin{array}{l}
X^i (\sigma,t) = x^i (\sigma,t) + x_f^i (\sigma,t) \,,
\\ { } \\
\partial_tX^i (\sigma,t) = v^i (\sigma,t) + v_f^i (\sigma,t) \,.
\end{array}
\label{slowfast}
\ee
The quantum field $x_f^i$ can be expressed as a sum over modes in
the $(\tau,\sigma)$ frame, provided the frequency cutoff, which
separates the fast modes from the slow ones, is chosen to reflect
the exponentially improving resolution in $\tau$,
\beq
\label{qmodeexp}
\lefteqn{
x_f^i(\sigma,t) = \sqrt{{\alpha'\over 2}}
\sum_{n=1}^\infty W(n+\frac{\epsilon}{\tau})
\bigl[\frac{c_n^i}{\sqrt{n}} e^{-in(\tau{+}\sigma)}
} \nonumber \\
& &\qquad\>
+\frac{\tilde{c}_n^i}{\sqrt{n}} e^{-in(\tau{-}\sigma)}
+ H.c. \bigr] \,,
\eeq
where $\epsilon$ is some constant.  The expansion for $v_f^i$
is similar.  For a filter function one could use
$W(n+\frac{\epsilon}{\tau})=\theta(n+\frac{\epsilon}{\tau})
=\theta(n-\epsilon \, e^{t/4M})$.  Then an asymptotic observer would
indeed include only modes of exponentially higher frequency in the
definition of the quantum part of the field.  In actual
calculations it is better to use a filter where the step of the theta
function is smeared a little, in order to avoid unphysical effects
associated with a sharp edge cutoff in two dimensions.

The long-wavelength field $x^i$ will evolve nontrivially in
asymptotic time due to the continual feeding in of modes from the
quantum fields.  This evolution is described by a simple
Langevin equation \cite{mpt}
\begin{equation}
\label{Langevin}
\partial_tx^i =  \eta^i \,,
\end{equation}
with noise correlator
\beq
\label{noisecorr}
\lefteqn{
\langle \eta^i(1) \, \eta^j(2) \rangle \simeq
\frac{\alpha'}{2}\, \frac{\delta^{ij}}{4 M} \,
\delta(t_1{-}t_2) \, \cos(\frac{\epsilon \Delta \sigma}{\tau})
} \nonumber \\
& &\qquad\qquad
\times\exp \bigl[-\frac{\beta^2}{4} \bigl(\frac{\epsilon \Delta
\sigma}{\tau}\bigr)^2 \bigr] \,.
\eeq
The time-dependence in the correlator assures us
that the noise $\eta^i$ is white, and the spatial dependence tells us
that wee bits of string of parameter length $\Delta \sigma \simeq
| \tau | / (\beta \epsilon)$ evolve independently with
time.\footnote{A description of strings in terms of discrete bits
has been considered by Giles and Thorn \cite{thorn} and by
Klebanov and Susskind \cite{klesus}.}

The Langevin equation in (\ref{Langevin}) describes a {\it branching
diffusion\/} process which thermalizes the string configuration.
At any given point of the string the value
of the slowly-varying field $x^i(\sigma, t)$ experiences a
Brownian motion which is essentially unaffected by anything
lying outside the correlation length
$\Delta \sigma \approx |\tau |/\beta \epsilon$.  This
resembles the independence of the scalar field evolution in different
Hubble domains in the case of chaotic inflation \cite{alam}.
The correlation length decreases exponentially with asymptotic time
and therefore the number of such independent bits of the string
increases exponentially:
\begin{equation}\label{number}
N \sim \frac{2\pi\beta\epsilon}{| \tau |} =
2\pi \beta\epsilon \, \exp(t/4M) \, .
\end{equation}
The diffusion of a given bit of string is governed by the
Fokker-Planck equation,
\begin{equation}
\partial_t P(x^i,t) = \frac{\varsigma^2}{2}
\delta^{ij}
\partial_i \partial_j P(x^i,t) \, .
\end{equation}
where $P(x^i,t)$ is the normalized probability of finding that
bit at $(x^i,t)$.  On the stretched horizon of a finite-mass
black hole, $P \rightarrow constant$ at late times.

The coefficient of diffusion can be read off the correlator of the
noise, $\varsigma ^2 = \frac{\alpha'}{8M}$, and
the mean square transverse position is given by
\begin{equation}
\label{xsqt}
\langle x^i x^i \rangle = 2 \varsigma^2 t
=\frac{\alpha'}{4M} \,  t \, .
\end{equation}
This is the transverse spread, linear in asymptotic time, pointed
out by Susskind \cite{susskind}.
The string will spread to cover the area
of the black hole horizon in a time
\begin{equation}
\label{tspread}
t_S\sim g^2 M^3 \,,
\end{equation}
where $M$ and $t$ are measured in Planck units and $g$ is the
string coupling strength.  This is a short time compared to
the black hole lifetime if the string is weakly coupled.

Another important timescale is that on which the volume density of
string at the center of the distribution becomes $O(1/g^2)$ in
string units.  At this point string interactions can no longer be
ignored and the above calculations, which are all at the level of
free string theory, no longer apply.  The proper thickness of the
stretched horizon remains of order one in string units while the
average area occupied increases linearly with time.  At the same
time the average number of string bits grows exponentially so the
time $t_I$ at which string interactions become important at the
center of the distribution of string bits is:
\begin{equation}
\label{tint}
t_I \sim M \log{(1/g^2)} \,.
\end{equation}
For a macroscopic black hole this timescale is very much shorter
than the spreading time $t_S$.  An attractive possibility, pointed
out in \cite{susskind}, is that string interactions prevent the
volume density from exceeding $O(1/g^2)$ and that the central
density will level off at that value.  This volume density
corresponds to an $O(1)$ area density of string bits on the stretched
horizon measured in Planck units, the value suggested by the
Bekenstein-Hawking entropy.  If string interactions prevent the
density from increasing beyond $O(1/g^2)$, this will
produce an outward pressure that spreads the string bits much more
rapidly than the diffusion of the free theory \cite{susskind2}.

We have presented evidence in support of the view that a stretched
horizon made out of strings has a key role to play in resolving the
black hole information paradox.  For technical reasons the
discussion was limited to strings approaching the horizon of
a classical black hole in the limit of infinite mass.  We cannot
at present establish, although we find it very plausible, that
the stringy stretched horizon re-emits the original information
encrypted in apparently thermal Hawking radiation.
Another important issue to address concerns the underlying causal
properties of string theory which allow the spread of information
across the stretched horizon \cite{causal}.

\vskip 8pt
I wish to thank the organizers for the opportunity to lecture at
the Spring School and for their hospitality at the ICTP Trieste.
I have benefited from discussions on black holes with many friends and
colleagues and I want especially to thank my collaborators:
A. Mezhlumian, A. Peet, J. Russo, A. Strominger, L. Susskind and
J. Uglum.  I also thank S. Giddings, A. Strominger, and J. Uglum for
helpful comments on the manuscript.


\begin{thebibliography}{9}
\frenchspacing

\bibitem{hawking1} S.W.~Hawking,
Commun.~Math.~Phys. {\bf 43} (1975) 199.

\bibitem{hawking2} S.W.~Hawking,
Phys.~Rev. {\bf D14} (1976) 2460.

\bibitem{hawell} See {\it e.g.} S.W.~Hawking and G.F.R. Ellis,
The Large Scale Structure of Spacetime, Cambridge University
Press, Cambridge, 1973.

\bibitem{hawking3} S.W.~Hawking,
Commun.~Math.~Phys. {\bf 87} (1982) 395.

\bibitem{ehns} J. Ellis, J.~Hagelin, D.V.~Nanopoulos, and
M.~Srednicki,
Nucl.~Phys. {\bf B241} (1984) 381.

\bibitem{bsp} T.~Banks, L.~Susskind, and M. Peskin,
Nucl. Phys. {\bf B244} (1984) 125.

\bibitem{gross} D.~Gross,
Nucl.~Phys. {\bf B236} (1984) 349.

\bibitem{strominger} A. Strominger,
{\it Unitary Rules for Black Hole Evaporation,}
hep-th 9410187, UCSBTH-94-34, October 1994.

\bibitem{page1} D.~Page,
Phys.~Rev.~Lett. {\bf 44} (1980) 301.

\bibitem{thooft1} G.~`t~Hooft,
Nucl.~Phys. {\bf B335} (1990) 138;
Phys.~Scr. {\bf T36} (1991) 247, and references therein.

\bibitem{susskind} L.~Susskind,
Phys.~Rev.~Lett. {\bf 71} (1993) 2367;
Phys.~Rev. {\bf D49} (1994) 6606.

\bibitem{acn} Y.~Aharonov, A.~Casher, and S.~Nussinov,
Phys.~Lett. {\bf B191} (1987) 51.

\bibitem{remnants} T.~Banks and M.~O'Loughlin,
Phys. Rev. {\bf D47} (1993) 540;
T. Banks, M. O'Loughlin, and A. Strominger,
Phys. Rev. {\bf D47} (1993) 4476;
A. Strominger and S. Trivedi,
Phys. Rev. {\bf D48} (1993) 5778;
S.B. Giddings,
Phys. Rev. {\bf D48} (1994) 947; {\it ibid.} 4078.

\bibitem{jpas} J. Polchinski and A.~Strominger,
{\it A Possible Resolution of the Black Hole
Information Puzzle},
UCSBTH-94-20, hep-th 9407008, July 1994.

\bibitem{cghs} C.G.~Callan, S.B.~Giddings, J.A.~Harvey, and
A.~Strominger,
Phys. Rev. {\bf D45} (1992) 1005.

\bibitem{jhas} J.A.~Harvey and A.~Strominger,
in the Proceedings of the TASI Summer School, June 3-28, 1992,
Boulder, Colorado, (World Scientific, 1993).

\bibitem{giddings} S.B.~Giddings,
in the Proceedings of the International Workshop on Theoretical
Physics, 6th Session, June 21-28, 1992, Erice, Italy,
(World Scientific, 1993).

\bibitem{gs1} S.B.~Giddings and A.~Strominger,
Phys. Rev. {\bf D46} (1992) 627.

\bibitem{bddo} T.~Banks, A.~Dabholkar, M.R.~Douglas, and
M.~O'Loughlin,
Phys. Rev. {\bf D45} (1992) 3607.

\bibitem{horowitz} G.~Horowitz, in
String Theory and Quantum Gravity '92, J.A. Harvey \etal\
eds., World Scientific (1993).

\bibitem{witten} E. Witten,
Phys. Rev. {\bf D44} (1991) 314.

\bibitem{rst1} J.G.~Russo, L.~Susskind, and L.~Thorlacius,
Phys. Lett. {\bf B292} (1992) 13.

\bibitem{chrful} S.M. Christensen and S.A. Fulling,
Phys. Rev. {\bf D15} (1977) 2088.

\bibitem{gidnel} S.B. Giddings and W.M. Nelson,
Phys. Rev. {\bf D46} (1992) 2486.

\bibitem{jacobson} For a recent discussion of this issue see
T. Jacobson, Phys. Rev. {\bf D48} (1993) 728.

\bibitem{bghs} B. Birnir, S.B.~Giddings, J.A.~Harvey, and
A.~Strominger,
Phys. Rev. {\bf D46} (1992) 638.

\bibitem{lslt1} L. Susskind and L. Thorlacius,
Nucl. Phys. {\bf B382} (1992) 123.

\bibitem{hawking4} S.W. Hawking,
Phys. Rev. Lett. {\bf 69} (1992) 406.

\bibitem{numerical} D.A. Lowe,
Phys. Rev. {\bf D47} (1993) 2446;
T. Piran and A. Strominger,
Phys. Rev {\bf D48} (1993) 4729.

\bibitem{bilcal} A. Bilal and C.G. Callan,
Nucl. Phys. {\bf B394} (1993) 73.

\bibitem{dealwis} S.P. de~Alwis,
Phys. Lett. {\bf B289} (1992) 278; {B300} (1993) 330;
Phys. Rev. {\bf D46} (1992) 5429.

\bibitem{rst2} J.G.~Russo, L.~Susskind, and L.~Thorlacius,
Phys. Rev. {\bf D46} (1992) 3444;
Phys. Rev. {\bf D47} (1993) 533.

\bibitem{strominger1} A. Strominger,
Phys. Rev. {\bf D48} (1993) 5769.

\bibitem{strtri} A. Strominger and S. Trivedi, unpublished.

\bibitem{stu} L. Susskind, L. Thorlacius, and J. Uglum,
Phys. Rev. {\bf D48} (1993) 3743.

\bibitem{chuver}T.D.~Chung and H.~Verlinde,
Nucl. Phys. {\bf B418} (1994) 305.

\bibitem{dasmuk} S.R.~Das and S.~Mukherji,
Phys. Rev. {\bf D50} (1994) 930.

\bibitem{aslt} A. Strominger and L. Thorlacius,
Phys. Rev. {\bf 50} (1994) 5177.

\bibitem{ssstt} N. Seiberg, S. Shenker, L. Susskind,
L. Thorlacius, and J. Tuttle, unpublished (1992).

\bibitem{mirrors} See e.g. N. Birrel and P. Davies
{\it Quantum Fields in Curved Space} (Cambridge University Press,
1982) and references therein;  R. Carlitz and S. Willey,
Phys. Rev. {\bf D36} (1987) 2327, 2336;
F. Wilczek, IAS preprint HEP-93/12, hep-th 9302096.

\bibitem{anddew} A. Anderson and B. DeWitt,
Found. Phys. {\bf 16} (1986) 91.

\bibitem{jrat} J.G. Russo and A.A. Tseytlin,
Nucl. Phys. {\bf B382} (1992) 259.

\bibitem{gidstr2} S.B.~Giddings and A.~Strominger,
Phys. Rev. {\bf D47} (1993) 2754.

\bibitem{burcha} T. Burwick and A. Chamseddine,
Nucl. Phys. {\bf B384} (1992) 411.

\bibitem{cgquant} J. Polchinski,
Nucl. Phys. {\bf B324} (1989) 123;
T. Banks, Physicalia Magazine, vol 12, Special Issue in
Honor of the 60th Birthday of R.Brout, Gent (1990);
S.R. Das, S. Naik, and S.R. Wadia,
Mod. Phys. Lett. {\bf A4} (1990) 1033;
S.R. Das, A. Dhar, and S.R. Wadia,
Mod. Phys. Lett. {\bf A5} (1990) 799;
T. Banks and J. Lykken,
Nucl. Phys. {\bf B331} (1990) 173;
A.A. Tseytlin,
Int. Jour. Mod. Phys. {\bf A5} (1990) 1833.

\bibitem{strominger2} A.~Strominger,
Phys. Rev. {\bf D46} (1992) 4396.

\bibitem{verl} E.~Verlinde and H.~Verlinde,
Nucl. Phys. {\bf B406} (1993) 43;
K.~Schoutens, E.~Verlinde, and H.~Verlinde,
Phys. Rev. {\bf D48} (1993) 2690.

\bibitem{cardy} J.L. Cardy, Nucl. Phys. {\bf 240}
(1984), 514.

\bibitem{lpst} D. Lowe, J. Polchinski, A. Strominger,
and L. Thorlacius, unpublished (1994).

\bibitem{olcnvf} O. Lechtenfeld and C. Nappi,
Phys. Lett. {\bf B288} (1992) 72;
V. Frolov, Phys. Rev. {\bf D46} (1992) 5383.

\bibitem{charged} T. Banks and M. O'Loughlin,
Phys. Rev. {\bf D48} (1993) 698;
D. Lowe and M. O'Loughlin,
Phys. Rev. {\bf D48} (1993) 3735;
A. Strominger and S. Trivedi,
Phys. Rev. {\bf D48} (1993) 5778.

\bibitem{maggiore} M. Maggiore,
Phys. Rev. {\bf D49} (1994) 2918;
Phys. Lett. {\bf B333} (1994) 39.

\bibitem{mathur}
S.D. Mathur,
{\it Black Holes Entropy and the Semiclassical Approximation,}
January 1994, hep-th 9404135;
E. Keski-Vakkuri, G. Lifschytz, S.D. Mathur, and M. Ortiz,
{\it Breakdown of the Semiclassical Approximation at the
Black Hole Horizon}, MIT-CTP-2341, hep-th 9408039,
July 1994;

\bibitem{mempar} K.S. Thorne, R.H. Price, and D.A.
MacDonald, Black Holes: The Membrane Paradigm,
Yale University Press, New Haven, CT, 1986, and
references therein.

\bibitem{shell} W.G. Unruh,
Phys. Rev. {\bf D14} (1976) 870.
Y. Choquet-Bruhat,
Ann. Inst. Henri Poincar\'e {\bf 8} (1968) 327;

\bibitem{lslt2} L. Susskind and L. Thorlacius,
Phys. Rev. {\bf D49} (1994) 966.

\bibitem{jphv} J. Preskill, private communication (1993);
H. Verlinde, private communication (1993).

\bibitem{page2} D. Page,
Phys. Rev. Lett. {\bf 71} (1993) 1291.

\bibitem{kks} L. Susskind,
Phys. Rev. {\bf D1} (1970) 1182;
M. Karliner, I. Klebanov, and L. Susskind,
Int. J. Mod. Phys. {\bf A3} (1988) 1981.

\bibitem{mpt} A. Mezhlumian, A. Peet, and L. Thorlacius,
Phys. Rev. {\bf D50} (1994) 2725.

\bibitem{thorn} R. Giles and C.B. Thorn,
Phys. Rev. {\bf D16} (1977) 366;
C.B. Thorn,
Phys. Rev. {\bf D17} (1978) 1073; {\bf D19} (1979) 639;
{\it Calculating the Rest Tension for a Polymer of String
Bits}, UFIFT-HEP-94-8, hep-th 9407169, July 1994.

\bibitem{klesus} I. Klebanov and L. Susskind,
Nucl. Phys. {\bf B309} (1988) 175.

\bibitem{alam} A.D. Linde and A. Mezhlumian,
Phys. Lett. {\bf B307} (1993) 25;
A.D. Linde, D.A. Linde, and A. Mezhlumian,
Phys. Rev. {\bf D49} (1994) 1783.

\bibitem{susskind2} L. Susskind,
{\it The World as a Hologram}, SU-ITP-94-33,
hep-th 9409089, September 1994.

\bibitem{causal} E. Martinec,
Class. Quant. Grav. {\bf 10} (1993) L187;
D.A. Lowe, Phys. Lett. {\bf B326} (1994) 223;
D.A. Lowe, L. Susskind, and J. Uglum,
Phys. Lett. {\bf B327} (1994) 226.

\end{thebibliography}
\end{document}